\documentclass[a4paper]{article}

\usepackage{epsfig}
\usepackage{color}
\usepackage{graphicx}
\usepackage{colortbl}

\definecolor{lg}{rgb}{0.2, 0.77, 1}
\definecolor{vlg}{rgb}{0.56, 0.87, 1}

\definecolor{lightgrey}{rgb}{0.9, 0.9, 1}
\definecolor{grey}{rgb}     {0.5, 0.5,0.5}
\definecolor{orange}{rgb}   {1,0.6,0.05}
\definecolor{brightorange}{rgb}{1,0.7,0.05}
\definecolor{lila}{rgb}     {0.8,0,0.5}
\definecolor{darkbrown}{rgb}{0.6,0.3,0}
\definecolor{brown}{rgb}    {0.8,0.4,0.}
\definecolor{darkblue}{rgb} {0.05,0,0.5}
\definecolor{colblue}{rgb}  {0,0.1,0.95}

\usepackage{amsmath,amsfonts,amssymb}
\usepackage{setspace} 

\newcommand{\la}[1]{}
\newcommand{\coout}[1]{}

\newcommand{\tab}[1]    {\texttt{\color{darkblue} #1}}
\newcommand{\col}[1]    {\texttt{\color{colblue} #1}}
\newcommand{\defext}[1] {\texttt{\color{darkbrown} #1}}
\newcommand{\defint}[1] {\texttt{\color{brown} #1}}
\newcommand{\nick}[1]   {\texttt{\color{grey} #1}}

 \usepackage{amsmath,amsfonts,amssymb}

\usepackage{epsfig}
\usepackage{graphicx}

\renewcommand{\headsep}{-1cm}
\renewcommand{\textheight}{25cm}

\begin{document}

\title{
\parbox{3.5cm}{\includegraphics[width=3cm]{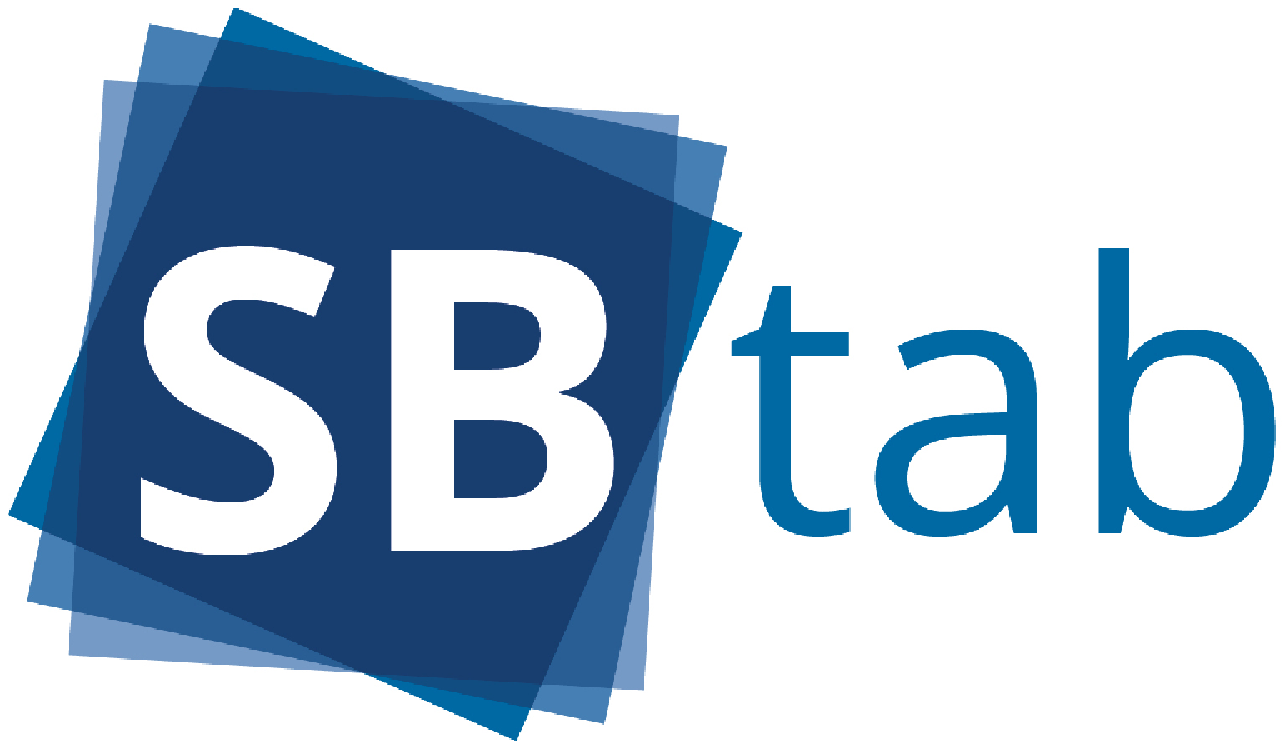}}
\parbox{10cm}{Conventions for structured   data tables  in\\ 
 Systems Biology -- SBtab version 0.9}
}

\author{Wolfram Liebermeister$^{1}$, Timo Lubitz$^{2}$, and Jens Hahn$^{2}$\\[5mm]
$^{1}$ Institut f\"ur Biochemie, Charit\'e - Universit\"atsmedizin Berlin\\ 
$^{2}$ Institut f\"ur Biophysik, Humboldt-Universit\"at zu Berlin}
\date{}

\maketitle

\begin{abstract}
  Data tables in the form of spreadsheets or delimited text files are
  the most utilised data format in Systems Biology. However, they are
  often not sufficiently structured and lack clear naming conventions
  that would be required for modelling. We propose the SBtab format as
  an attempt to establish an easy-to-use table format that is both
  flexible and clearly structured. It comprises defined table types
  for different kinds of data; syntax rules for usage of names,
  shortnames, and database identifiers used for annotation; and
  standardised formulae for reaction stoichiometries.  Predefined
  table types can be used to define biochemical network models and the
  biochemical constants therein.  The user can also define own table
  types, adjusting SBtab to other types of data. Software code, tools,
  and further information can be found at www.sbtab.net.
\end{abstract}

\section{Introduction}

Spreadsheets and delimited text tables are the most utilised data
formats in Systems Biology. They are easy to use and can hold various
types of data. Tables can not only store omics data, but also
metabolic network models described by lists of biochemical reactions.
However, when tables are exchanged within scientific collaborations,
modellers usually prefer tables that can be processed automatically,
and the flexibility of spreadsheets can become a disadvantage. If
table structures and nomenclature vary from case to case, parsing
becomes laborious and new files require new parsers. Furthermore,
different naming conventions -- for instance, for biochemical
compounds -- make it hard to combine data, for instance metabolic
network models and omics data produced by different researchers.
Therefore, rules for structuring tables and for consistent naming and
annotations can make tables much more useful as exchange formats in
Systems Biology collaborations and for usage in software tools. The
SBtab format comprises a set of conventions for data tables that are
supposed to make tables easier and safer to work with. Let us start
with a couple of examples.  Then we continue with a more formal
specification of SBtab version 0.9.

\paragraph{Example 1: Structure of a metabolic network model}
A stoichiometric metabolic model can be defined by a list of
biochemical reaction formulae, specifying the substrates, products,
and their stoichiometric coefficients. Such reactions can be listed
in a single column of a spreadsheet, and additional
information may be provided: each reaction can have a number or
identifier (defined only within the model) and can be linked to an
entry in the database KEGG Reaction \cite{KEGG}. Furthermore,
reactions may be catalysed by enzymes, which  relates them  to
certain genes. All information could be stored in the following
table:

\begin{center} {\tt {\footnotesize
      \begin{tabular}{|l|l|l|l|}
        \hline
         Reaction & Sum formula & KEGG ID & Gene symbol\\ \hline
        R1 & ATP + F6P <=> ADP + F16P &R00658 & pfk \\ \hline
        R2 & F16P + H2O <=> F6P + Pi &R01015 & fbp \\ \hline
      \end{tabular}}\ \\[5mm]
  }
\end{center}

where \texttt{ATP}, \texttt{F6P}, \texttt{ADP}, \texttt{F16P},
\texttt{H2O}, and \texttt{Pi} are shortnames for metabolites to be
used in the model. Although the information is complete and
unambiguous, the parser still has to recognise that the columns
\texttt{Sum formula} and \texttt{KEGG ID} contain reaction formulae
and identifiers in certain formats. If the column names and the syntax
of the reaction formulae vary from table to table (e.g.~\texttt{<->}
is used instead of \texttt{<=>}), parsing becomes tedious. In the
SBtab format, the table would look a little more complicated, but is
easy to parse automatically:

\begin{center} {\tt \footnotesize {
      \begin{tabular}{|l|l|l|l|}
        \hline
        \cellcolor{lg}\tab{!!SBtab} & \cellcolor{lg}\tab{TableName='Ex 1 - Reaction'}& \cellcolor{lg}\tab{TableType='Reaction'}& \cellcolor{lg}\\
        \hline
        \cellcolor{vlg}\col{!Reaction} & \cellcolor{vlg}\col{!SumFormula} &\cellcolor{vlg}\col{!Identifiers}:\defext{kegg.reaction} & \cellcolor{vlg}\col{!Gene:Symbol}\\ \hline
        \nick{R1} & \nick{ATP} + \nick{F6P} <=> \nick{ADP} + \nick{F16P} &R00658 & pfk \\ \hline
        \nick{R2} & \nick{F16P} + \nick{H2O} <=> \nick{F6P} + \nick{Pi} &R01015 & fbp \\ \hline
      \end{tabular}}\ \\[5mm]
  }
\end{center}

In this table, elements highlighted by colours have special meanings
(the colours themselves are just used in this text and are not part of
the SBtab format). The SBtab table differs from the original table in
several ways: the first line (starting with \col{!!}) declares that
the table is an SBtab table of the type \tab{Reaction} and must
therefore satisfy syntax rules for this table type. The following line
contains the column headers. They start with the \col{!} character,
emphasising that they were not chosen \emph{ad hoc} by the user, but stem
from a controlled vocabulary. The predefined column headers do not
contain whitespaces. The header \texttt{KEGG ID} has been replaced by
the term \col{!Identifiers:}\defext{kegg.reaction}.  This may look
complicated, but it allows parsers to retrieve further data from
databases in a stable way\footnote{The expression
  \defext{kegg.reaction} is defined by the MIRIAM resources and used
  within SBtab. The URL of the KEGG database, defining the
  identifiers, may change in the future; however, KEGG's Miriam ID
  (provided by the the MIRIAM resources web service
  \cite{laibe2007miriam}) is guaranteed to remain stable in time.}.
The syntax of the reaction formulae is also uniquely defined. In
particular, the shortnames of metabolites must not contain any
whitespaces or special characters, which simplifies parsing and makes
them suitable as variable names for computer models.  The meaning of
these shortnames can be defined by providing standardised names or
database identifiers in a second table of type \tab{Compound}.  The
compound shortnames will then serve as keys to rows of this table.

\begin{center} {\tt \footnotesize
    \begin{tabular}{|l|l|l|}
\hline
    \cellcolor{lg}\tab{!!SBtab} & \cellcolor{lg}\tab{TableName='Ex 2 - Compound'} & \cellcolor{lg}\tab{TableType='Compound'} \\
      \hline
\cellcolor{vlg}\col{!Compound}& \cellcolor{vlg}\col{!Name} & \cellcolor{vlg}\col{!Identifiers:}\defext{kegg.compound}\\
      \hline
      \nick{F6P} & Fructose 6-phosphate & C05345 \\ \hline
      \nick{ATP} & ATP & C00002 \\ \hline
      \nick{ADP} & ADP & C00008 \\ \hline
      \nick{F16P}& Fructose 1,6-bisphosphate & C00354 \\ \hline
      \nick{H2O} & Water & C00001 \\ \hline
      \nick{Pi} & Inorganic phosphate & C00009 \\ \hline
      \nick{PEP} & Phosphoenolpyruvate & C00074 \\ \hline
      \nick{AMP} & AMP & C00020 \\
      \hline
\end{tabular}
}
\end{center}

Both tables together form an SBtab document describing a model. In
practice, they can be stored as separate files, as sheets of a
spreadsheet file, or within a single table. The following example
contains all necessary information to build a stoichiometric model in
the SBML (Systems Biology Markup Language) format \cite{hfsb:03}:

{\tt \footnotesize
\begin{center}
\begin{tabular}{|l|l|l|l|l|}
\hline
\cellcolor{lg}\tab{!!SBtab} & \cellcolor{lg}\tab{TableName='Ex 3 - Reaction'} & \cellcolor{lg}\tab{TableType='Reaction'} & \cellcolor{lg}\\
\hline
\cellcolor{vlg}\col{!Reaction} & \cellcolor{vlg}\col{!SumFormula} & \cellcolor{vlg}\col{!Identifiers:}\defext{kegg.reaction} & \cellcolor{vlg}\col{!SBML:reaction:id} \\ \hline
\nick{R1} & \nick{ATP} + \nick{F6P} <=> \nick{ADP} + \nick{F16P} &R00658 &r1 \\ \hline
\nick{R2} & \nick{F16P} + \nick{H2O} <=> \nick{F6P} + \nick{Pi} &R01015 &r2 \\ \hline
\cellcolor{lg}\tab{!!SBtab} & \cellcolor{lg}\tab{TableName='Ex 3 - Compound'} & \cellcolor{lg}\tab{TableType='Compound'} & \cellcolor{lg}\\
\hline
\cellcolor{vlg}\col{!Compound}& \cellcolor{vlg}\col{!Name} & \cellcolor{vlg}\col{!Identifiers:}\defext{kegg.compound}& \cellcolor{vlg}\col{!SBML:species:id}\\
\hline
\nick{F6P} & Fructose 6-phosphate & C05345 &f6p \\ \hline
\nick{ATP} & ATP & C00002 &atp \\ \hline
\nick{ADP} & ADP & C00008 &adp \\ \hline
... & ... & ... & ... \\
\hline
\end{tabular}
\end{center}
}

Here, we have added new identifiers (in the columns \col{SBML:reaction:id}
and \col{SBML:species:id}) for  \tab{Reaction}
and \tab{Compound} entries to be used in SBML. Such extra names could be 
necessary if the original shortnames do not comply with SBML's rules
for element identifiers.

\paragraph{Example 2: Table of kinetic constants} In a second example, we
specify numerical parameters, for example kinetic constants and
metabolite concentrations that appear in a kinetic model. Each
quantity can be related to a compound (e.g.~a concentration), to a
reaction (e.g.~an equilibrium constant), or to several biological
elements (e.g.~to an enzyme and a compound, in the case of Michaelis-Menten
constants). As in the previous example, these elements can be specified
by unique identifiers, e.g.~KEGG compound or reaction identifiers.
Furthermore, each quantity has a value and a physical unit. In the
SBtab format, we arrange this information in a table of type
\tab{Quantity}. Each row contains all information about
one of the quantities:

{\tt \footnotesize
  \begin{center}
    \begin{tabular}{|l|l|l|l|l|l|l|l|l|l|l|l|l|l|l|ll}
      \hline
      \cellcolor{lg}\tab{!!SBtab} & \cellcolor{lg}\tab{TableName='Ex 4 - Quantity'} & \cellcolor{lg}\tab{TableType='Quantity'} & \cellcolor{lg}& \cellcolor{lg}& \cellcolor{lg}\\\hline
      \cellcolor{vlg}\col{!Quantity} & \cellcolor{vlg}\col{!QuantityType} & \cellcolor{vlg}{\tiny \col{!Reaction:Identifiers:}\defext{kegg.reaction}} & \cellcolor{vlg}{\tiny \col{!Compound:Identifiers:}\defext{kegg.compound}} & \cellcolor{vlg}\defint{!Value} & \cellcolor{vlg}\col{!Unit} \\ \hline
      \nick{keq\_R1}& \defext{equilibrium constant} & R01061 & & 0.156 & {\tiny \defext{dimensionless}}\\ \hline
      \nick{kmc\_R1\_C1}& \defext{Michaelis constant} & R01061 & C00003 & 0.96 & \defext{mM}\\ \hline
      \nick{kic\_R1\_C1}& \defext{inhibition constant} & R01070 & C00111 & 0.13 & \defext{mM}\\ \hline
      \nick{con\_C1}& \defext{concentration} & & C00118 & 0.203 & \defext{mM}\\ \hline
      ... & ... & ... & ... & ... & ... \\ \hline
    \end{tabular}
  \end{center}
}

The first two columns specify a name and a type for each quantity.
The quantity types (\defext{substrate catalytic rate constant},
\defext{equilibrium constant} etc.) are not chosen \emph{ad hoc}, but
stem from the Systems Biology Ontology (SBO) \cite{cjkw:11}. This
ensures a unique spelling and allows software to retrieve definitions
and further information from the SBO web services. The biological
elements (in this case, reactions, compounds, or both) are specified
in the following two columns by unique identifiers from the KEGG
database. Columns with human-readable names, or identifiers from other
databases, could be added. Unnecessary fields remain empty. The column
name \defint{Value} -- like some other mathematical terms -- is
defined for SBtab (arbitrary values in this example). Unit names are
defined as in SBML (see below).  If the table is used together with a
metabolic model, we can use compound and reaction identifiers from the
model instead of the Identifiers.org annotations \cite{junl:12}. In
this case, the table would read:

\begin{center} {\tt \footnotesize
    \begin{tabular}{|l|l|l|l|l|l|l|l|l|l|l|l|l|l|ll}
      \hline
      \cellcolor{lg}\tab{!!SBtab} & \cellcolor{lg}\tab{TableName='Ex 5 - MyData'} & \cellcolor{lg}\tab{TableType='Quantity'} & \cellcolor{lg}&\cellcolor{lg}&\cellcolor{lg}\\
      \hline
      \cellcolor{vlg}\col{!Quantity} &       \cellcolor{vlg}\col{!QuantityType} & \cellcolor{vlg}\col{!SBML:reaction:id} & \cellcolor{vlg}\col{!SBML:species:id} & \cellcolor{vlg}\defint{!Value} & \cellcolor{vlg}\col{!Unit} \\ \hline
 \nick{MyData\_1} &      \defext{equilibrium constant} & r1 & & 0.156 & \defext{dimensionless}\\ \hline
 \nick{MyData\_2} &      \defext{Michaelis constant} & r1 & atp & 0.96 & \defext{mM}\\ \hline
 \nick{MyData\_3} &      \defext{inhibition constant} & r1 & atp & 0.13 & \defext{mM}\\ \hline
 \nick{MyData\_4} &      \defext{concentration} & & atp & 1.5 & \defext{mM}\\ \hline
 ... &  ... & ... & ... & ... & ... \\ \hline
\end{tabular}
}
\end{center}

This table, together with a stoichiometric model and a choice of
standardised rate laws (like the modular rate laws \cite{liuk:10})
completely defines a kinetic metabolic model.

\paragraph{Example 3: A table with metabolome data} As a last example,
let us consider a table with metabolome time series data. For the sake
of simplicity, only two metabolites (rows) and measured samples
(columns) are shown:
\begin{center} {\tt \footnotesize {
  \begin{tabular}{|l|l|l|l|l|}
    \hline
    \cellcolor{lg}\tab{!!SBtab} & \cellcolor{lg}\tab{TableType='QuantityMatrix'}& \cellcolor{lg}\tab{TableName='Ex 6 - Metabolomics data'}& \cellcolor{lg} \tab{UniqueKey='False'} & \cellcolor{lg} \\
    \hline
    \cellcolor{vlg}\col{!Compound}& \cellcolor{vlg}\col{!Identifiers:obo.chebi}& t = 0 s &  t = 0.5 s & ..\\ \hline
Glucose  & CHEBI:17234 & 1.1 & 1.2 & ..\\ \hline
Fructose & CHEBI:15824 & 1.4 & 0.9 & ..\\ \hline
.. & .. & .. & .. & ..\\ \hline
      \end{tabular}}\ \\[5mm]
  }
\end{center}
Tables of this sort can be also be used for other kinds of omics data.
In this example, the headers of data columns (e.g., \texttt{t = 0 s})
do not follow a specific syntax and contain relevant information (time
point and time unit). We shall see below how such information can be
provided in SBtab in a more structured manner.

In the following sections, we introduce the general SBtab rules
(specification for SBtab version 0.9), as well as formats and
conventions for different types of use (see Section
\ref{chapter2}). It defines a list of table types (see Section
\ref{chapter3}) and explains the syntax of reaction \la{and
  regulation} formulae in the SBtab format (see Section
\ref{chapter4}).  Finally, the specification references the available
online tools for the handling of SBtab files (see Section
\ref{chapter5}) and includes an overview of all available SBtab table
types in appendix \ref{appendixA}. Appendix \ref{appendixB} lists
controlled vocabularies and database resources recommended to be used
within SBtab.

\section{Overview of the SBtab format}
\label{chapter2}

\subsection{Basic conventions}

SBtab comprises a list of conventions about the structure,
nomenclature, syntax, and annotations in tables describing biochemical
network models, kinetic parameters, and dynamic data. It contains
\begin{enumerate}
\item General rules for the \textbf{structure of tables} and the \textbf{syntax} used in table fields.
\item Defined \textbf{table types} for different kinds of
  information, each with possible \textbf{columns} with defined names
  and data types (see Table \ref{tab:tabletypes}; An overview of all
  predefined table types and their possible columns is given in the
  appendix).
\item A \textbf{syntax for biochemical element annotations} pointing to databases
  or ontologies.
\item Rules for usage of \textbf{names}, \textbf{shortnames}, and \textbf{database identifiers}
  used for annotation.
\item \textbf{Naming rules for biochemical quantities} to specify the quantities,
  physical units, and mathematical terms (like \defint{Mean} for mean values).
\item A syntax for \textbf{reaction sum formulae}. \la{simple kinetics, and regulation.}
\item A mechanism for  \textbf{extending the format} by declaring new column or table types.
\end{enumerate}
While the general rules apply to all kinds of data, the current
version of SBtab is tailored for describing the structure of
biochemical network models and the biochemical quantities therein.
This is reflected by the table types defined
in Table \ref{tab:tabletypes}.
\la{; (iii) omics data.} 

\textbf{Colour highlighting and predefined terms} In the examples
shown in this text, predefined SBtab entries are highlighted in
colours.  This is just for convenience and is not a part of the SBtab
format.  \tab{Table types} and \col{Column types} defined by the SBtab
format are listed in Table \ref{tab:tabletypes}. \nick{Shortnames} can
be chosen \emph{ad hoc} by the user; each of them needs to be defined
by a table row.  Shortnames have to be unique and consistent within a
document, but may differ between documents. \defint{Reserved names}
are predefined in SBtab for recurrent mathematical expressions like
``mean value''.  \defext{Official names}, like the names used for
databases, are defined by some other authority. \texttt{Free text} and
other text including database IDs, numerical values, mathematical
brackets, and operators is written in black.

\subsection{SBtab tables and  SBtab documents}

\textbf{General table structure} An SBtab document consists of one or
several tables that refer to a common model or related data sets.  All
tables must use a common list of shortnames. For instance, a
\tab{Compound}\, table contains the column \col{!Compound}, and the
elements from this column define compound shortnames to be used in the
other tables.  Several tables in a document may have the same type,
but their table names (attribute \tab{TableName}) must be unique.

\textbf{Declaration row containing the table attributes} The top left field
contains the table header, starting with \tab{!!SBtab} and followed by
the table attributes in the syntax \emph{attribute
  name}\tab{=\'}\emph{attribute value}\tab{\'}, separated by
whitespaces. Mandatory attributes are \tab{TableType} and
\tab{TableName}.

\textbf{Column headers and definition table} The second row contains
the column headers. Columns whose headers start with a \col{!} are
treated as SBtab columns and must adhere to the SBtab rules.  Other
columns can contain arbitrary content.  SBtab has a number of
predefined table types that can hold different kinds of data. Each
table type has a number of mandatory or optional columns with specific
properties. An overview is given below and in the appendix.  However,
users can also define their own table types and corresponding
columns. This definition must be provided by the user in the form of a
special \tab{Definition} table (as described below).

\textbf{Column with unique keys} By default, any SBtab table must start with a
column matching its table type (e.g., a table of type \tab{Quantity}
must start with a column \col{Quantity}) and containing shortnames
that serve as unique identifiers for the table elements. If a table
does not have such a column with unique key, this should be marked by
setting the attribute \col{UniqueKey='False'} in the declaration
row of the table. The attribute is set to \col{True} by default.

\textbf{Completeness} To interpret the contents of a single table, other
tables (e.g.~describing shortnames) may be required. If a table does
not require any other tables, we call it ``complete''. A document is
complete if all names are defined, i.e.~no unspecified information is
required to interpret its contents. If a single table or a document
are incomplete, the undefined names have to be known by the software,
and an exchange with other software tools is likely to
fail. \la{Different IDs or names for the same elements have to agree,
  otherwise the parser will throw an error.} If a table or document
contains two elements, and there is no explicit information implying
that they describe the same things, it is assumed that they describe
different things.

\textbf{Conventions for spreadsheet files} To ensure consistency
between spreadsheet files, we propose a number of rules for good practice:
\begin{itemize}
\item \textbf{UTF8 encoding} If possible, the UTF8 encoding should be
  chosen.
\item \textbf{Documents}
  In character-separated text files (.csv or .tsv), a document can either
  be stored in several files with the filenames
  \emph{basename}\texttt{\_}\emph{tablename}.\emph{extension}, or
  tables are concatenated vertically, each preceded by a declaration
  row (starting with \tab{!!}), and stored in a single table file.
\item \textbf{Delimiters in .csv or .tsv files} In character-separated
  files, irrespective of the extension (.csv or .tsv), it is assumed
  by default that the delimiters are tabulators. However, other delimiters
  (comma or semicolon) are accepted by the parser as well.
\item \textbf{Special characters} If table cells contain special
  characters that are also used as cell delimiters (e.g.~commas), the
  file must be provided in a form that excludes ambiguities (e.g.~in
  the case of a comma-separated table containing commas with its
  fields, all cells must additionally be marked by quotation marks
  (\texttt{".."}).
\end{itemize}

 \textbf{Filenames} The
  SBtab format as such does not impose any restrictions on filenames,
  nor does it require a specific filename extension.  SBtab files
  stored as excel sheets, for instance, will have the extension
  \texttt{.xls}. However, the SBtab online tools (and the python programs
  behind it) have a certain convention for filenames and filename
  extensions.  When an SBtab document is exported to several delimited
  text files, the filenames will be chosen according to the scheme
  \texttt{[SBTAB DOCUMENT NAME]\_[TABLE TYPE].csv} or, in case of ambiguities
  \texttt{[SBTAB DOCUMENT NAME]\_[TABLE TYPE]\_[TABLE NAME].csv}.

 \textbf{Filename extensions}  Regarding filename extensions, the python implementation of SBtab
 supports comma-separated and tab-separated tables, as well as excel
 spreadsheet files (xls). By default, the python code exports tab-separated
 files and uses the filename extension \texttt{.csv}.  This is a convention
 supported by LibreOffice, but may lead to conflicts in other
 cases. Some tools require extensions like \texttt{.tab} (excel) or \texttt{.tsv}
 (e.g., the formatting option in github). We do not use \texttt{.tsv} in this
 case, because this is supported neither by excel nor by
 LibreOffice. In case of conflicts, users may have to simply rename
 their files.  When importing a table, the code tries to determine whether
 commas or tabs are used as delimiters. When using commas as
 delimiters, users have to make sure that no commas are used
 elsewhere in the table (or that all elements are given in double
 quotes).

 \begin{table}[t!]
   \begin{center}
     \begin{tabular}{|l|l|l|}
       \hline
       Name & Contents & Usage \\
       \hline
       \tab{Compound} & Names, IDs, properties of compounds & model structure \\
       \tab{Enzyme} & Names, properties of enzymes   & model structure \\
       \tab{Protein} & Names, properties of proteins & model structure \\
       \tab{Gene} & Names, properties of genes       & model structure \\
       \tab{Regulator} & Names, properties of gene regulators & model structure \\
       \tab{Compartment} & Names and IDs of compartments & model structure \\
       \tab{Reaction} & Chemical reactions & model structure \\
       \tab{Quantity} & Individual data for model parameters & quantitative data \\
       \tab{QuantityMatrix}&  Data matrices & quantitative data \\
       \tab{Relation} & Relations between different compounds & model structure\\
       \tab{Definition} & Define custom column types, etc. & customise SBtab \\
       \hline
     \end{tabular}
   \end{center}
   \caption{Overview of  table types predefined in SBtab. \label{tab:tabletypes}}
 \end{table}

\subsection{Names of biochemical elements}

\paragraph{Names and identifiers of model elements}
In the following, compounds, enzymes, genes, genetic regulators, and
compartments will be called ``biochemical entities''. ``Biochemical
elements'' comprises, in addition, reactions and biochemical
quantities. Biochemical elements can be described by shortnames,
official names, or database identifiers (IDs). The shortnames have to
be declared within the SBtab document and have to satisfy syntactic
rules. Each table starts with a column of the same name, containing
the shortnames. Shortnames, the arbitrary element names used in a data
set or model, must be unique, i.e.~declared only once in a document;
they must start with a letter and may not contain spaces or
the special characters ``:'', ``.''.  In columns containing database
IDs, the column name (\col{!Identifiers:}\emph{Identifiers}) specifies the
database by a name (to be used in column names, IDs etc.) and an URI.
We suggest to use preferably the databases listed in the Miriam file
(see Table \ref{tab:databases}).  Sometimes, elements may
be characterised redundantly: e.g.~the reaction catalysed by an
enzyme, given in an \tab{Enzyme} table, can be given by both shortname
and database ID. In case of conflict, the information derived from the
shortname (i.e.~the database ID listed in the \tab{Reaction} table)
has higher priority.

\la{Other databases can be declared
  in a table of type \tab{AnnotationResource}.}

\paragraph{Naming and specification of biological entities}
Tables of the types \tab{Compound}, \tab{Enzyme}, \tab{Gene},
\tab{Regulator}, or \tab{Compartment} are called ``entity tables''.
The biochemical meaning of the entities can be
 declared by different columns:
\begin{itemize}
\item \col{!Name} contains official names (it is good practice to use
  names from the suggested databases).  Several names can be listed in
  one field, separated by ``$|$''.  To declare from which database a
  name has been taken, the name can also be written as
  \emph{DB}:\emph{name}.
\item \col{!Identifiers:}\emph{Identifiers} contains IDs from a
  specified database.  Annotations with database IDs follow the scheme
  defined by Identifiers.org \cite{junl:12} (data collection and ID). \la{????
    Databases listed in the MIRIAM resources \cite{MIRIAM} are denoted
    in formulae by the name as given by the MIRIAM URN string.}
\end{itemize}

\paragraph{Localised compounds}
If a compound, enzyme, or genetic regulator is localised in a
compartment, the corresponding localised entity can be denoted by
\emph{compound}\texttt{[}\emph{compartment}\texttt{]} with square
brackets, where \emph{compound} and \emph{compartment} are the
shortnames or IDs of the compound and the compartment used in the
model. If a model contains several compartments, tools should treat
the first compartment in the \tab{Compartment} table as the standard
compartment. The standard compartment will be used by default for all
compounds that are not explicitly assigned to compartments.

\subsection{Annotating biochemical elements  with database  identifiers}

\la{There are two types of identifier columns:
  1.} 

Biochemical elements  are annotated with database IDs listed in special
identifier columns. An \col{Identifiers} column contains annotations from one web
resource, at most one annotation per element, and without qualifiers.
The column item and the referenced ID are assumed to be linked by an
``is'' relationship (and not, for instance, ``version of'', which can
exist in SBML annotations). A table can contain several \col{Identifiers}
columns, which must refer to different data resources.

{\tt
  \begin{center} \footnotesize
    \begin{tabular}{|l|l|l|l|l|}
      \hline
      \cellcolor{lg}\tab{!!SBtab} & \cellcolor{lg}\tab{TableName='Ex 7 - Compound'} & \cellcolor{lg}\tab{TableType='Compound'} & \cellcolor{lg}\\ \hline
      \cellcolor{vlg}\col{!Compound} & \cellcolor{vlg}\col{!Identifiers:}\defext{obo.chebi}& \cellcolor{vlg}\col{!Identifiers:}\defext{kegg.compound}& \cellcolor{vlg}... \\
      \hline
      \nick{water} & CHEBI:15377 & C00001 & ...\\\hline
      \nick{ATP} & CHEBI:15422 & C00002 & ...\\\hline
      \nick{phosphate} & CHEBI:18367 & & ...\\\hline
    \end{tabular}
  \end{center}
}

To translate an element like \texttt{CHEBI:16865} into a valid
Identifiers.org URI, \texttt{http://identifiers.org/} is concatenated
with the data collection mentioned after \col{!Identifiers:} in the
header (e.g.~\defext{obo.chebi}) and with the column item, separated
by a slash\footnote{The elements from the column have to be translated
  into a URN-encoded form (as described in the URN specification): for
  instance, the colon in the identifier \texttt{CHEBI:16865} has to be
  replaced by the string ``\texttt{\%3A}'' to create the URN
  \texttt{obo.chebi:CHEBI\%3A16865}.}. For instance, the first
annotation entry in the table above would be resolved to 
\texttt{http://identifiers.org/obo.chebi/CHEBI:15377}. 

\subsection{Syntax for reaction formulae}
\label{chapter4}

Chemical reactions can be described by reaction formulae (column
\col{!SumFormula} in table \tab{Reaction}; specifying the reactants,
their stoichiometric coefficients, and possibly their localisation).
The reaction arrow is denoted by \texttt{<=>}. Stoichiometric
coefficients refer to substance amounts, not concentrations (this
matters in the case of transport reactions).  Stoichiometric
coefficients of 1 are omitted; general stoichiometric coefficients,
given by letters (e.g.~\texttt{n}) are not allowed. If possible, the
reaction formula should represent the actual stoichiometries
experienced by the enzyme (i.e.  \texttt{\nick{A} <=> 2 \nick{B}}
rather than \texttt{0.5 \nick{A} <=> \nick{B}}).  Substrates and
products are given by shortnames, which must be defined in a
\tab{Compound} table.  The order of substrates and the order of
products are arbitrary; however, comparison of formulae is eased by
using an alphabetical order. The localisation in compartments can be
denoted as follows:

\begin{itemize}
\item Reaction in the default compartment: \texttt{\nick{A} + 2 \nick{B} <=> \nick{C} + \nick{D}}
\item Transport reaction: \texttt{\nick{A}[\nick{comp1}] + 2 \nick{B}[\nick{comp1}] <=> \nick{C}[\nick{comp2}] + \nick{D}[\nick{comp2}]}
\end{itemize}

In the example, \nick{A}, \nick{B}, \nick{C}, and \nick{D} are
compound shortnames, and \nick{comp1} and \nick{comp2} are compartment
shortnames. The reversibility of reactions is not given by the sum formula, 
but by an extra column \col{!IsReversible} in the \tab{Reaction} table.

\section{Overview of predefined table types}
\label{chapter3}

SBtab predefines a number of table types with specific properties.  An
overview is given in Table \ref{tab:tabletypes}. The table types
\tab{Compound}, \tab{Enzyme}, \tab{Gene}, \tab{Regulator},
\tab{Compartment}, and \tab{Reaction} describe model structures, the
table types \tab{Quantity}, \tab{QuantityMatrix}, and \tab{Relation} \la{,
  \tab{OmicsDataRow}, and \tab{OmicsDataColumn}} are used for
quantitative data.

\subsection{Tables for biochemical network structures}

As in  example  1 (in the introduction section), biochemical
networks consist of biochemical entities (e.g.~metabolites or proteins)
and reactions or interactions between them. The tables describing these
entities (table types \tab{Reaction}, \tab{Compound}, \tab{Compartment},
\tab{Enzyme}, \tab{Regulator}, and \tab{Gene}) have to satisfy the following
rules.

\begin{itemize}

\item \textbf{Entities} In tables describing biochemical entities
  (\tab{Compound}, \tab{Enzyme}, \tab{Gene}, \tab{Regulator},
  \tab{Compartment}), each row has to contain (i) a shortname as the
  primary key (in the column \col{!Compound}, \col{!Enzyme}, etc.) and
  (ii) at least one entry specifying the entity, like \col{!Name} or
  \col{!Identifiers:}\emph{DB}. If a column shares the type of the
  table (e.g.~a Compound column in a Compound table), it can be
  considered a primary key, that is, its elements should be unique and
  it should appear as the first column in the table. Optional columns
  - which may depend on the kinds of entities - are listed in Table
  \ref{tab:columnsentities}.

\item \textbf{Reactions}
A \tab{Reaction} table lists chemical reactions, possibly with
information about the corresponding enzymes, their kinetic laws, and
their genetic regulation. It must contain at least one of the
following columns: \col{!SumFormula}, \col{!Identifiers:}\emph{DB}; optional
columns are listed in Table \ref{tab:columnsreactions}. For an
example, see example 1 in the introduction.

\item
\textbf{Enzymes, genes, and regulators}
The connection between chemical reactions, the enzymes catalysing the
reactions, and the genes coding for the enzymes can be complicated,
but in many cases, there is a one-to-one relationship. In SBtab, there
are different ways to express this relationship. Information about
enzymes or genes and their regulation can be stored in a
\tab{Reaction} table if there is a one-to-one relationship between
reactions, enzymes, and possibly genes. Otherwise, it is stored in
separate tables \tab{Enzyme} and \tab{Gene} and the tables are
interlinked \emph{via} the columns
\col{!Enzyme} (in table \tab{Reaction}) and
\col{!Gene} (in table \tab{Enzyme}) or
\col{!TargetReaction} (in an \tab{Enzyme}
table) and \col{!GeneProduct} (in a \tab{Gene}
table).

\end{itemize}

\subsection{Table type \tab{Quantity} for biochemical parameters}

Numerical data (e.g.~for time series or kinetic parameters) can be
stored in tables and be linked to model elements via the latters'
shortnames. There are two \la{four} different table types for
numerical data.  Tables of type \tab{Quantity} describe individual
physical or biochemical quantities, for instance, kinetic parameters
in a network model. These quantities can be linked to one entity, one
reaction or enzyme, or both. If a quantity table contains several
values for the same quantity, they appear in separate rows (for
possible descriptions of provenance, see Table
\ref{tab:columnsalltables}).

Tables of type \tab{Quantity} describe single physical or biochemical
quantities (e.g.~individual kinetic constants).
A quantity is defined by a type, a unit, possibly
biochemical entities to which it refers, possibly a localisation, and
possibly experimental or physical conditions. The columns contain the
defining properties (e.g. unit, conditions, etc.) and their values.
Quantities can refer to a compound, an enzyme or reaction, or a
combination of them. For instance, a concentration refers to a
substance, while a $k^{\rm M}$ value refers to a metabolite and an
enzyme. If there is a one-to-one relationship between reactions and
enzymes, the $k^{\rm M}$ value can also be assigned to a
compound/reaction pair or a compound/enzyme pair. Let us consider
again example 2:\\

{\tt \scriptsize
\begin{tabular}{|l|l|l|l|l|l|l|l|l|l|l|l|l|l|l|}
\hline
\cellcolor{lg}\tab{!!SBtab}	& \cellcolor{lg}\tab{TableName='Ex 8 - Quantity'} & \cellcolor{lg}\tab{TableType='Quantity'} & \cellcolor{lg}& \cellcolor{lg}& \cellcolor{lg}\\ \hline
\cellcolor{vlg}\tab{!Quantity} & \cellcolor{vlg}\col{!QuantityType} & \cellcolor{vlg}{\tiny \col{!Reaction:Identifiers:}\defext{kegg.reaction}} & \cellcolor{vlg}{\tiny \col{!Compound:Identifiers:}\defext{kegg.compound}} & \cellcolor{vlg}\defint{!Value} & \cellcolor{vlg}\col{!Unit} \\ \hline
\nick{keq\_R1} & \defext{equilibrium constant} & R01061 & & 0.0984 & dimensionless \\ \hline
\nick{kmc\_R1\_C1} & \defext{Michaelis constant} & R01061 & C00003 & 0.96 & mM\\ \hline
\nick{kic\_R1\_C1} & \defext{inhibition constant} & R01070 & C00111 & 0.13 & mM\\ \hline
\nick{con\_C1} & \defext{concentration} & & C00118 & 0.203 & mM\\ \hline
\end{tabular}
}

To specify the parameters of a model, we refer to \col{Reaction} and
\col{Compound} elements by shortnames rather than by resource IDs. In
this form, the above example becomes

\begin{center}
 {\tt \footnotesize
    \begin{tabular}{|l|l|l|l|l|l|}
\hline
    \cellcolor{lg}\tab{!!SBtab} & \cellcolor{lg}\tab{TableName='Ex 9 - Quantity'} & \cellcolor{lg}\tab{TableType='Quantity'} & \cellcolor{lg}& \cellcolor{lg}& \cellcolor{lg}\\
      \hline
      \cellcolor{vlg}\col{!Quantity} & \cellcolor{vlg}\col{!SBO:Identifiers:}\defext{obo.sbo} & \cellcolor{vlg}\col{!Reaction} & \cellcolor{vlg}\col{!Compound} & \cellcolor{vlg}\defint{!Value} & \cellcolor{vlg}\col{!Unit} \\ \hline
      \nick{kcrf\_R1} & SBO:0000320 & R1 & & 200.0 & 1/s\\ \hline
      \nick{keq\_R1} & SBO:0000281 & R1 & & 0.0984 & dimensionless \\ \hline
      \nick{kmc\_R1\_C1} & SBO:0000027 & R1 & C1 & 0.96 & mM \\ \hline
      \nick{kic\_R1\_C2} & SBO:0000261 & R1 & C2 & 0.13 & mM\\ \hline
      \nick{con\_C3} & SBO:0000196 & & C3 & 0.203 & mM\\
      \hline
\end{tabular}
}
\end{center}

This example shows that quantity types can be specified by
identifiers from the Systems Biology Ontology (SBO) in a column
\col{!SBO:Identifiers:}\defext{obo.sbo}.

A \tab{Quantity} table can also store state-dependent
quantities like concentrations, expression levels, or fluxes, like in
the following example.

\begin{center} {\tt \footnotesize
    \begin{tabular}{|l|l|l|l|l|l|l|}
      \hline
    \cellcolor{lg}\tab{!!SBtab} & \cellcolor{lg}\tab{TableName='Ex 12 - Quantity'} & \cellcolor{lg}\tab{TableType='Quantity'}& \cellcolor{lg}& \cellcolor{lg}\\
\hline
      \cellcolor{vlg}\col{!Quantity} & \cellcolor{vlg}\col{!Compound} & \cellcolor{vlg}\col{!Condition} & \cellcolor{vlg}\col{!SBO:concentration} & \cellcolor{vlg}\col{!Unit} \\ \hline
      \nick{con\_C1\_wt} & C1 & wildtype & 0.2 & mM\\ \hline
      \nick{con\_C2\_wt} & C2 & wildtype & 1 & mM\\\hline
      \nick{con\_C3\_wt} & C3 & wildtype & 0.1 & mM\\\hline
      \nick{con\_C1\_mu} & C1 & mutant   & 0.1 & mM\\\hline
      \nick{con\_C2\_mu} & C2 & mutant   & 0.5 & mM\\\hline
      \nick{con\_C3\_mu} & C3 & mutant   & 0.1 & mM\\\hline
\end{tabular}
}
\end{center}

\subsection{Table type \tab{QuantityMatrix} for data matrices}

Biological data often have the form of matrices.  As an example,
consider a small 2$\times$2 matrix containing metabolite
concentrations for two time points and two metabolites. It can be
expressed by the following SBtab table.

\begin{center} {\tt \footnotesize {
      \begin{tabular}{|l|l|l|l|}
        \hline
        \cellcolor{lg}\tab{!!SBtab} & \cellcolor{lg}\tab{TableType='QuantityMatrix'}& \cellcolor{lg}\tab{TableName='Ex 13 - Metabolomics data'} \tab{UniqueKey='False'}\\
        \hline
       \cellcolor{vlg}\col{!Time} &  Glucose &  Fructose \\ \hline
       0.0  & 1.1 &          1.4\\ \hline
       0.5  & 1.2 &          0.9\\ \hline
      \end{tabular}}\ \\[5mm]
  }
\end{center}
The headers of the data columns are not defined headers starting with
 ``!'', but simple strings. Therefore, they are not formally
 controlled by SBtab. Annotating these columns, e.g., by adding ChEBI
 Identifiers to specify the metabolites, is not directly
 possible. Moreover, the time points have no keys to which other
 tables could refer. An alternative solution looks as follows:

\begin{center} {\tt \footnotesize {
      \begin{tabular}{|l|l|l|l|}
        \hline
        \cellcolor{lg}\tab{!!SBtab} & \cellcolor{lg}\tab{TableType='QuantityMatrix'}& \cellcolor{lg}\tab{TableName='Ex 14 - Metabolomics data'}& \cellcolor{lg}\tab{UniqueKey='False'}\\
        \hline
        \cellcolor{vlg}\col{!TimePoint}& \cellcolor{vlg}\col{!Time} & \cellcolor{vlg}\col{>Measurement:Glucose} &  \cellcolor{vlg}\col{>Measurement:Fructose} \\ \hline
T0 &        0.0  & 1.1 &          1.4\\ \hline
T1 &        0.5  & 1.2 &          0.9\\ \hline
      \end{tabular}}\ \\[5mm]
  }
\end{center}
Here, the column headers are controlled and point to rows of another table with table name ``Measurement'', in which the ChEBI Identifers are given:
\begin{center} {\tt \footnotesize {
  \begin{tabular}{|l|l|l|l|}
    \hline
    \cellcolor{lg}\tab{!!SBtab} & \cellcolor{lg}\tab{TableType='Quantity'} & \cellcolor{lg} \tab{TableName='Ex 15 - Measurement'} & \cellcolor{lg}\tab{UniqueKey='False'}\\
    \hline
    \cellcolor{vlg}\col{!Compound}& \cellcolor{vlg}\col{!Identifiers:obo.chebi} & \cellcolor{vlg}\col{!QuantityType} & \cellcolor{vlg}\col{!Unit}\\ \hline
Glucose  & CHEBI:17234 & concentration & mM\\ \hline
Fructose & CHEBI:15824 & concentration & mM\\ \hline
      \end{tabular}}\ \\[5mm]
  }
\end{center}

Now let us consider data tables in which time points are represented
by columns. A similar scheme can be used in this case. The first,
simple version would read:
\begin{center} {\tt \footnotesize {
  \begin{tabular}{|l|l|l|l|}
    \hline
    \cellcolor{lg}\tab{!!SBtab} & \cellcolor{lg}\tab{TableType='QuantityMatrix'}& \cellcolor{lg}\tab{TableName='Ex 16 - Metabolomics data'}& \cellcolor{lg} \tab{UniqueKey='False'}\\
    \hline
    \cellcolor{vlg}\col{!Compound}& \cellcolor{vlg}\col{!Identifiers:obo.chebi}& t = 0 s &  t = 0.5 s\\ \hline
Glucose  & CHEBI:17234 & 1.1 & 1.2 \\ \hline
Fructose & CHEBI:15824 & 1.4 & 0.9 \\ \hline
      \end{tabular}}\ \\[5mm]
  }
\end{center}

Here, it would obviously be good to store time point and time unit  separately instead of merging them in the column header. This can be realised as follows: 

\begin{center} {\tt \footnotesize {
  \begin{tabular}{|l|l|l|l|}
    \hline
    \cellcolor{lg}\tab{!!SBtab} & \cellcolor{lg}\tab{TableType='QuantityMatrix'}& \cellcolor{lg}\tab{TableName='Ex 17 - Metabolomics data'} & \cellcolor{lg}\tab{UniqueKey='False'}\\
    \hline
    \cellcolor{vlg}\col{!Compound} & \cellcolor{vlg}\col{!Identifiers:obo.chebi}&  \cellcolor{vlg} \col{>TimePoint:t0} &  \cellcolor{vlg} \col{>TimePoint:t1}\\ \hline
Glucose   & CHEBI:17234 & 1.1 &          1.2 \\ \hline
Fructose  & CHEBI:15824 & 1.4 &          0.9 \\ \hline
      \end{tabular}}\ \\[5mm]
  }
\end{center}
with an extra table
\begin{center} {\tt \footnotesize {
  \begin{tabular}{|l|l|l|}
    \hline
    \cellcolor{lg}\tab{!!SBtab} & \cellcolor{lg}\tab{TableType='Quantity'} &  \cellcolor{lg}\tab{TableName='Ex 18 - TimePoint' UniqueKey='False'}\\
    \hline
 \cellcolor{vlg}\col{!TimePoint} & \cellcolor{vlg}\col{!Time}& \cellcolor{vlg}\col{!Unit}\\ \hline
t0 & 0   & s\\ \hline
t1 & 0.5 & s\\ \hline
      \end{tabular}}\ \\[5mm]
  }
\end{center}

\subsection{The table type \tab{Relation} for pairwise relations}

The table type \tab{Relation} is used to define pairwise links between
objects. Each link (``relationship'') can have a type and a numerical
value. A \tab{Relation} table can, for instance, be used to define a
directed graph (by listing the edges between nodes of one type) or a
gene regulatory network (by listing the actions of transcription
factors on gene promoters). In particular, \tab{Relation} tables can
be used to link SBtab elements between tables and, thus, to create
SBtab documents that have the form of a relational database.

\begin{center} {\tt \footnotesize
    \begin{tabular}{|l|l|l|l|l|l|l|}
      \hline
    \cellcolor{lg}\tab{!!SBtab} & \cellcolor{lg}\tab{TableName='Ex 19 - LittleGraph'} & \cellcolor{lg}\tab{TableType='Relation'}& \cellcolor{lg}\tab{UniqueKey='False'}\\
\hline
      \cellcolor{vlg}\col{!From} & \cellcolor{vlg}\col{!To} & \cellcolor{vlg}\col{!Relation} & \cellcolor{vlg}\col{!Value} \\ \hline
      \nick{A} & \nick{A} & regulates & 1\\ \hline
      \nick{A} & \nick{B} & regulates & -1\\\hline
      \nick{B} & \nick{A} & regulates & 1\\\hline
      \nick{B} & \nick{C} & regulates & 2\\\hline
      \nick{C} & \nick{D} & regulates & 1\\\hline
\end{tabular}
}
\end{center}

\subsection{Table type \tab{Definition} for customising the SBtab format}

Users can define their own table types and corresponding columns. For
usage in the online tools or in the python code, this definition can
be provided by the user in the form of a special \tab{Definition}
table.  The default table (containing the predefined table and column
types) is available on the SBtab website. Note that, when using a new
Definition table, the predefined Definition table will be completely
overridden, so any tables and columns to be used (also the predefined
ones) must be listed in the new table. \coout{@Timo: this should also me
  mentioned on the website in the place where definition tables can be
  uploaded.} The typical format of a \tab{Definition} table is shown
below.
			
{\tt \footnotesize
  \begin{center} 
    \begin{tabular}{|l|l|l|l|l|l|}
      \hline 
      \cellcolor{lg}\tab{!!SBtab} & \cellcolor{lg}\tab{TableType='Definition'} & \cellcolor{lg}\tab{TableName='Ex 20 - Def'} & \cellcolor{lg} & \cellcolor{lg}\\ \hline
      \cellcolor{vlg}\col{!Component}& \cellcolor{vlg}\col{!ComponentType}& \cellcolor{vlg}\col{!IsPartOf} &  \cellcolor{vlg}\col{!Format} & \cellcolor{vlg}\col{!Description}\\
      \hline
      SBML:reaction:id    & Column & \nick{Reaction} & \defint{String} & SBML ID of reaction \\\hline
      SumFormula          & Column & \nick{Reaction} & \defint{String} & Reaction sum formula \\\hline
      Enzyme              & Column & \nick{Reaction} & \defint{String} & Enzyme catalysing the reaction \\\hline
      \nick{...} & ... & ... & ... & .. \\\hline
    \end{tabular}
  \end{center}
}

The \col{Format} column defines which type of entries a column can
contain.  Possibilities are \defint{String}, \defint{Shortname} (name
of SBtab element, as defined in one of the SBtab tables),
\defint{Number} (integer or float in usual formats, or complex numbers
like \texttt{1 + i 3}), or \defint{Boolean} (with possible values \defint{True} and \defint{False}, or alternatively 1 and 0). More specific string
formats (e.g., for reaction sum formulae) are currently not formally
defined, but can be mentioned in the \col{Description} column.

\section{Conversion between SBtab and SBML} 
\label{conversion}
SBML (Systems Biology Markup Language) models can be converted into
SBtab documents and vice versa. Depending on the content of the SBML
model, the SBtab files can comprise table types \tab{Reaction}, \tab{Compound},
\tab{Compartment}, \tab{Quantity}, \tab{Events}, and \tab{Rules}. 
Likewise, these SBtab table types can be converted into an SBML
(Level 2, Version 4) model. The conversion to SBML, however, requires
at least either a \tab{Reaction} or \tab{Compound} SBtab.

The conversion from an SBML model file to
SBtab translates the structural and temporal information of the model
into corresponding SBtab table files. The (i)
\tab{Reaction} SBtab contains a list of the reactions of the SBML
file, including their sum formula, kinetic laws, irreversibility,
annotations, and more. Note that the SBML modifiers of a reaction (e.g. enzymes)
cannot be identified as inhibitor or stimulator if they are not assigned an SBO Term
within the SBML code. If this is not the case, they will only be exported to SBtab
as modifiers without regulatory information. All species from the model can be found in
the (ii) \tab{Compound} SBtab. Their location, charge, annotations, and more
are provided in the SBtab. Analogously, a (iii) \tab{Compartment} holds
all structural information of the cellular compartments.
The (iv) \tab{Quantity} SBtab file lists all parameters that are
part of the model. Also their numerical values and units will be
provided. The parameters can appear as either local or global
variables in the SBML code; this information will be transferred to
SBtab as well. (v) \tab{Events} can be an important part of SBML models; they
indicate e.g. concentration changes or stress applications at certain
time points. They too are translated into an SBtab file. Finally, (vi)
\tab{rules} are exported from SBML to an SBtab Rule table. Rules can comprise
assignment rules, algebraic rules, and rate rules. Rule formulas and
units are part of the conversion as well.

In the conversion from SBtab to SBML, \col{Compound} entries in SBtab correspond to
\texttt{species} elements in SBML.  By default, the unique keys in the
\col{Compound} and \col{Reaction} SBtab are used as
\texttt{id} attributes of the SBML elements. If SBML IDs are directly
specified within SBtab (in the columns \col{SBML:reaction:id},
\col{SBML:species:id}, \col{SBML:parameter:id},
\col{SBML:reaction:parameter:id}, etc), these will be used instead.
Rate laws from the SBML code are stored in SBtab as strings within a
\col{KineticLaw} column. Note that the rate laws are not checked for
their validity. It is up to the user to assure the correctness of the
rate laws. If they are erroneous, this leads to invalid SBML output. An
automatic parser of rate laws including checks of validity is planned for
future versions of SBtab. \col{Regulator} entries in SBtab correspond
to \texttt{modifier} elements in SBML; multiple regulators can be
described by a regulation formula (in the \col{Regulator} column):
regulators are separated by a ``$|$'' symbol, while the sign of
regulation can be denoted by + or -. For an enzyme allosterically
inhibited by ATP and activated by ADP, the formula reads
\texttt{-ATP|+Pyruvate} or \texttt{ATP|ADP} where inhibition and
activition remain unspecified. Also rate rules and assignment rules
can be converted from SBtab to SBML. Note that, just like for kinetic rate laws,
these rules do not underlie constraints of validity. It is up to the user to
ensure their correctness before conversion to SBML. Finally, SBtab is able
to provide lists of events for the SBML file. This includes the event assignments,
triggers, delays, and more. For all aforementioned SBML elements, annotations are
automatically translated from the SBtab to the SBML file, if they adhere to the correct
syntax. 

The entries of \tab{Quantity} tables can
be inserted into SBML models or be extracted from them. By default,
SBtab quantities referring to a reaction will become local reaction
parameters in SBML, while other quantities become global
parameters. The element of the \col{!Quantity} column will be used as
SBML element ID unless it is overridden by the (optional) column
\col{!SBML:parameter:id} (for global parameters) or
\col{!SBML:reaction:parameter:id} (for local reaction parameters).
\la{Boolean value? Describe in TABLES} Naming conventions for kinetic
constants are given in \cite{liuk:10}, supplementary material Table
A.5. Quantities that describe initial species amounts, initial species
concentrations, or compartment sizes will be translated into the
corresponding SBML element attributes.

There are still limitations to the conversion of SBtab and SBML.
So far, the conversion does not include element notes and function 
definitions. These issues are planned to be solved in future versions of SBtab.

\section{SBtab tools}
\label{chapter5}

To simplify the usage of SBtab, we provide several
online tools at \texttt{www.sbtab.net}.\\

\parbox[b]{9cm}{ 
\textbf{1. Online validator for SBtab files.} The online validator
tool checks whether SBtab files (in .csv or .xls format) adhere to the
SBtab conventions introduced in this manuscript.  If a problem is
identified by the validator, an instruction on how to fix the problem
is provided. The validation is based on the SBtab table definitions
found in the \tab{Definition} table.
}\hspace{5mm}
\parbox[b]{5cm}{
\includegraphics[width=5cm]{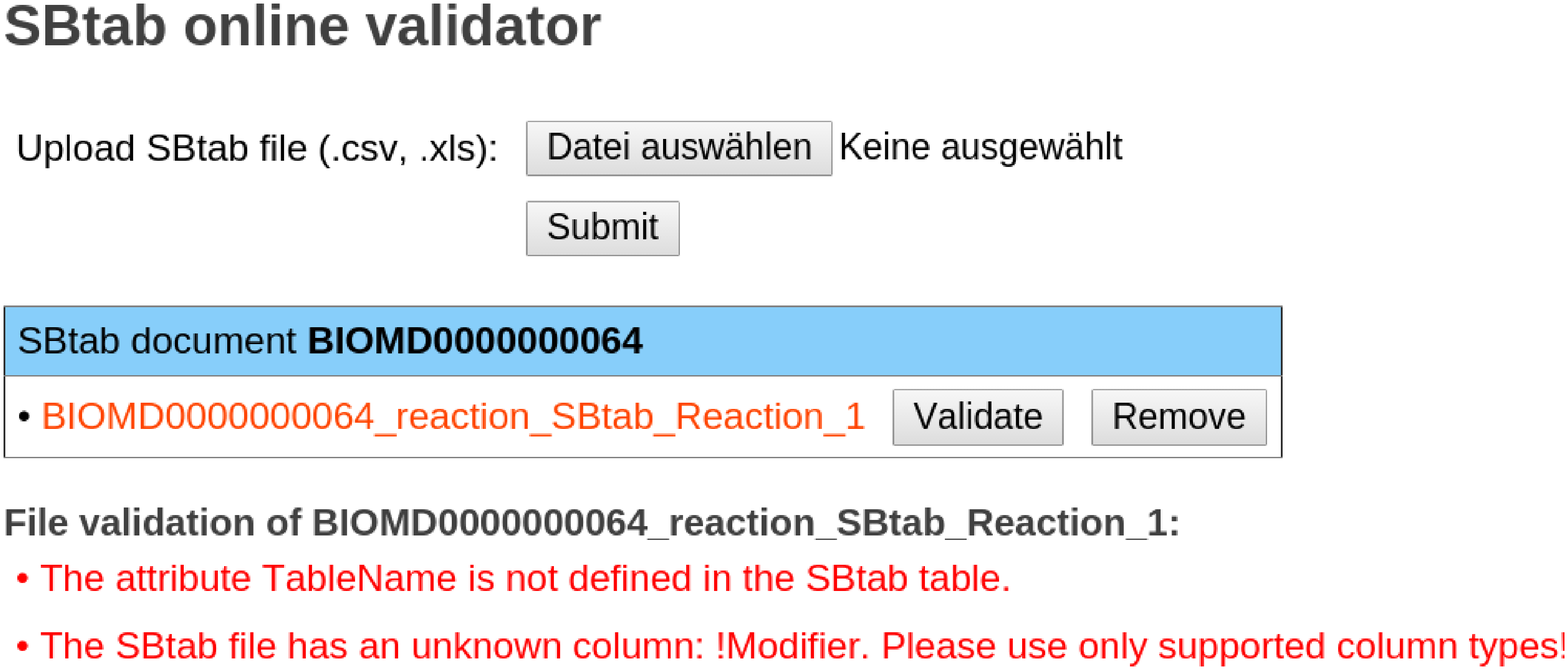}
}

\ \\
\parbox[b]{9cm}{ 
 \textbf{2. Online SBtab $\leftrightarrow$ SBML converter} The
  online conversion tool can create SBtab files from SBML models and
  vice versa. For the conversion from SBtab to SBML, it has to be
  assured that at least an SBtab  table of type \tab{Reaction}  or \tab{Compound}
  is provided. As
  additional information, the following SBtab table types can be used
  for the conversion to SBML:\tab{Compartment}, \tab{Quantity}, \tab{Events}, and \tab{Rules}.
  All information comprised in these SBtab tables
  can be converted to the SBML structure, as long as they are
  adhering to the correct syntax. Therefore, it is recommended
  to validate the SBtab files with the online validator before
  recruiting them for a conversion to SBML. The generated SBtab
  files can be displayed online as HTML tables. If annotations are
  correctly provided, they will link to the web resource.
  For the conversion it is recommended to use SBML Level 2, Version 4,
  or higher. The details on the conversions can be read in Chapter \ref{conversion}.
}\hspace{5mm}
\parbox[b]{5cm}{
\includegraphics[width=5cm]{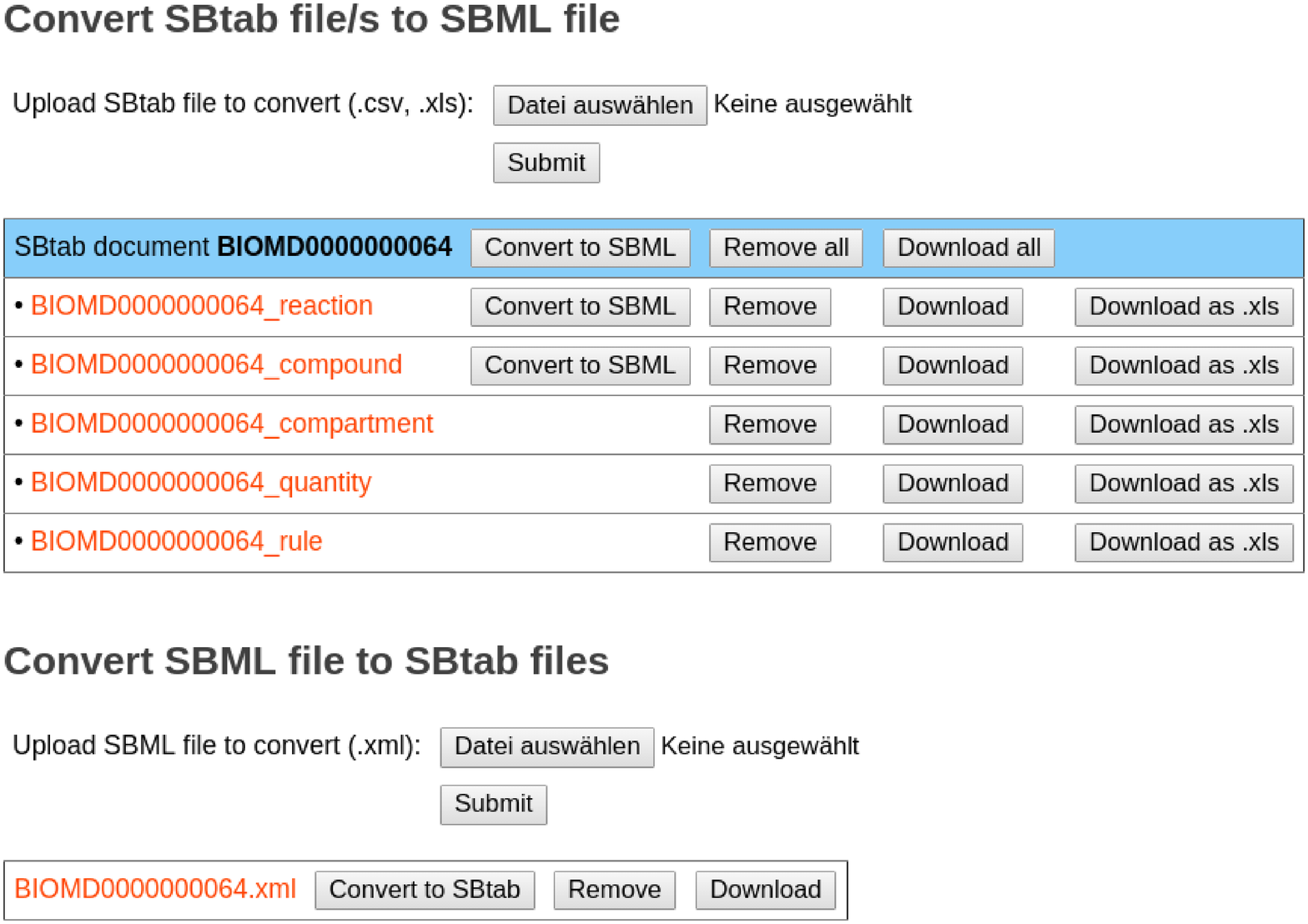}\\[2.5cm]
}

\ \\
\parbox[b]{9cm}{ 
 \textbf{3. MS Excel Add-in} The described validator and converter
functions can also be attained with an add-in for Microsoft Excel. It 
can be retrieved from the SBtab Github Repository and installed with a
Windows Installer Package. The prerequisites for the installation of the
add-in are (i) Windows Vista or higher, (ii) Microsoft Office 2010 or higher,
(iii) Microsoft .NET Framework 4.5 (full) or higher, and Microsoft Visual
Studio 2010 Tools for Office Runtime (VSTO). The latter two can be downloaded
directly from Microsoft.\\[.5cm]
}\hspace{5mm}
\parbox[b]{5cm}{
\includegraphics[width=5cm]{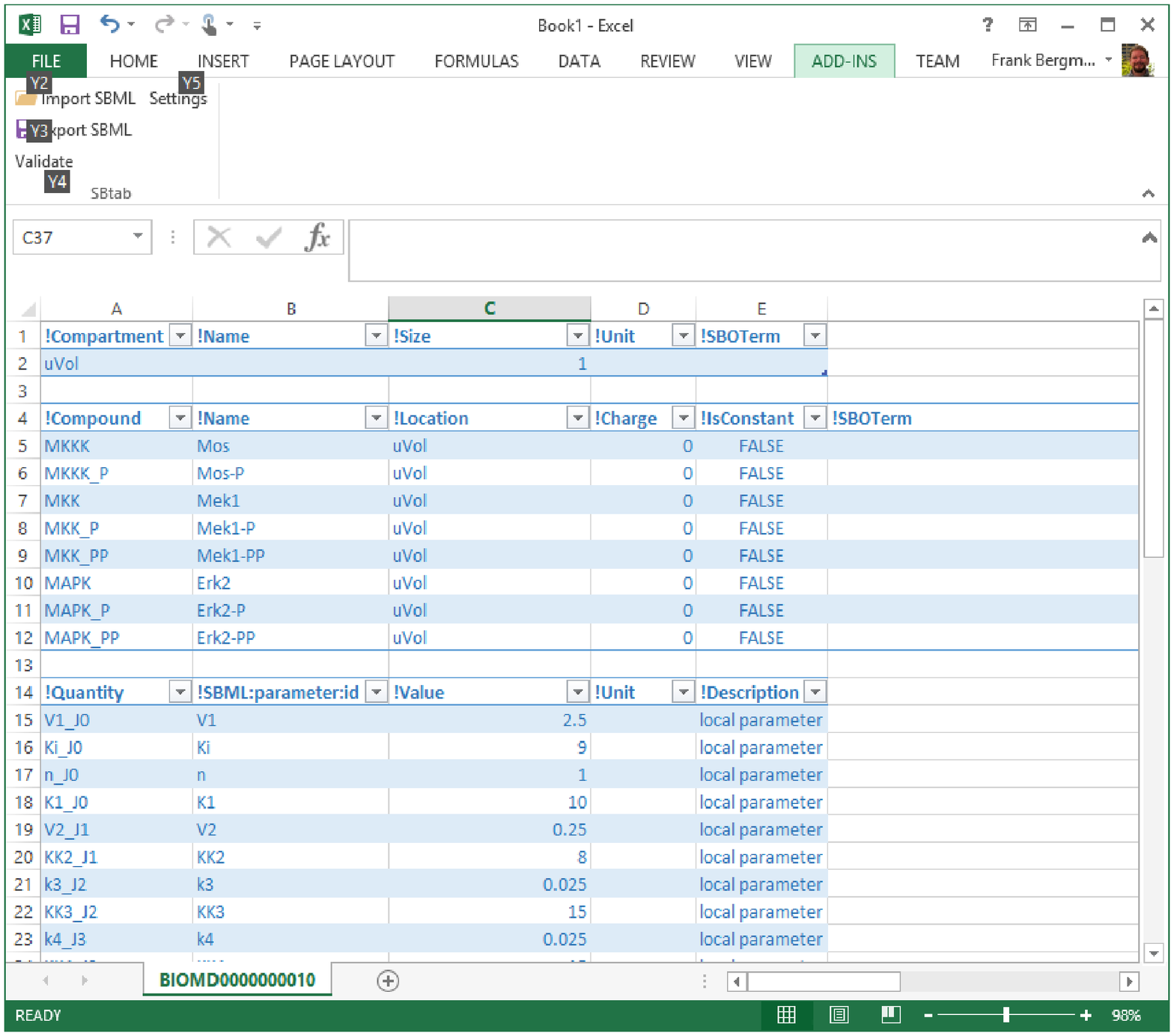}
}\ \\

 \textbf{Python parser for SBtab files.} In addition,  we provide a SBtab parser
 written in Python. It uses the Python package \texttt{tablib} to
 import SBtab files and provides different functions for editing the
 data and for directly accessing them.  These features are important
 for the embedding of the SBtab file parser into software
 projects. The common operations for manipulating SBtab files contain:
\begin{enumerate}
\item Extracting characteristic table information (type, name, etc.).
\item Addition of rows and columns to the SBtab table.
\item Editing and export of the table content in rows, columns, and
  single entries. An export as a Python dictionary is also possible,
  to ensure easy access to the data for python programmers.
\item Switching of columns and rows in the table (matrix
  transposition). As some data are stored conveniently in
  transposed spreadsheets, some tables need to be transposed to have
  better access to its content.
\item Duplicate SBtab objects.
\item Writing SBtab files to the hard disk.
\end{enumerate}

\section*{Acknowledgements}

The authors thank Dagmar Waltemath, Hans-Michael Kaltenbach, Dirk
Wiesenthal, Jannis Uhlendorf, Anne Goelzer, J\"org B\"uscher, Avi
Flamholz, Elad Noor, Edda Klipp, Frank Bergmann, Phillipp Schmidt, and
Matthias K\"onig for contributing to this proposal. This work was
funded by the German Research Foundation (LI 1676/2-1), the European
Commission (projects BaSysBio and UNICELLSYS), and the German Federal
Ministry of Education and Research (project OncoPath).

\bibliographystyle{unsrt}
\bibliography{/home/wolfram/latex/bibtex/biology}

\begin{thebibliography}{1}

\bibitem{KEGG}
M.~Kanehisa, S.~Goto, S.~Kawashima S, and A.~Nakaya.
\newblock The {KEGG} databases at genomenet.
\newblock {\em Nucleic Acids Research}, 30:42--46, 2002.

\bibitem{laibe2007miriam}
C.~Laibe and N.~Le~Nov{\`e}re.
\newblock {MIRIAM Resources: tools to generate and resolve robust
  cross-references in Systems Biology}.
\newblock {\em BMC Systems Biology}, 1(1):58, 2007.

\bibitem{hfsb:03}
M.~Hucka, A.~Finney, H.M. Sauro, H.~Bolouri, J.C. Doyle, H.~Kitano, A.P. Arkin,
  B.J. Bornstein, D.~Bray, A.~Cornish-Bowden, A.A. Cuellar, S.~Dronov, E.D.
  Gilles, M.~Ginkel, V.~Gor, I.I. Goryanin, W.J. Hedley, T.J. Hodgman, J.H.
  Hofmeyr, P.J. Hunter, N.S. Juty, J.L. Kasberger, A.~Kremling, U.~Kummer,
  N.~Le Nov\`ere, L.M. Loew, D.~Lucio, P.~Mendes, E.~Minch, E.D. Mjolsness,
  Y.~Nakayama, M.R. Nelson, P.F. Nielsen, T.~Sakurada T~J.C. Schaff, B.E.
  Shapiro, T.S. Shimizu, H.D. Spence, J.~Stelling, K.~Takahashi, M.~Tomita,
  J.~Wagner, J.~Wang, and the SBML~Forum.
\newblock {The Systems Biology Markup Language (SBML): A} medium for
  representation and exchange of biochemical network models.
\newblock {\em Bioinformatics}, 19(4):524--531, 2003.

\bibitem{cjkw:11}
M.~Courtot et~al.
\newblock Controlled vocabularies and semantics in systems biology.
\newblock {\em Molecular Systems Biology}, 7(543), 2011.

\bibitem{junl:12}
N.~Juty, N.~Le Nov\`ere, and C.~Laibe.
\newblock Identifiers.org and {MIRIAM Registry}: community resources to provide
  persistent identification.
\newblock {\em Nucleic Acids Research}, 40:D580--D586, 2012.

\bibitem{liuk:10}
W.~Liebermeister, J.~Uhlendorf, and E.~Klipp.
\newblock Modular rate laws for enzymatic reactions: thermodynamics,
  elasticities, and implementation.
\newblock {\em Bioinformatics}, 26(12):1528--1534, 2010.

\bibitem{GO}
The Gene~Ontology Consortium.
\newblock Gene ontology: tool for the unification of biology.
\newblock {\em Nature Genetics}, 25:25--29, 2000.

\bibitem{Novere:05}
N.~Le Nov\`ere, A.~Finney, M.~Hucka, U.S. Bhalla, F.~Campagne,
  J.~Collado-Vides, E.J. Crampin, M.~Halstead, E.~Klipp, P.~Mendes, P.~Nielsen,
  H.~Sauro, B.~Shapiro, J.L. Snoep, H.D. Spence, and B.L. Wanner.
\newblock {Minimum information requested in the annotation of biochemical
  models (MIRIAM)}.
\newblock {\em Nat Biotech.}, 23(12):1509--1515, Dec 2005.

\end{thebibliography}

\clearpage

\begin{appendix}

\section{Summary of SBtab rules}

We summarise the most important conventions implemented by  SBtab:
\begin{itemize}
\item \textbf{Shortnames} Model elements (e.g. compounds) are referred
  to by shortnames, which are defined in the corresponding table
  (e.g. \tab{Compound} for compounds) \la{or in the table \tab{Name}
    (for any kind of elements, properties, etc.)}. Shortnames must be
  unique within an SBtab document.  The first column of each table
  shares the name of the table type (e.g.~column \col{!Compound} in
  table type \tab{Compound}) and contains the shortnames, which serve
  as primary keys for this table and must therefore be unique. If a
  table does not contain such a unique key column, this must be
  declared by setting the table attribute \col{UniqueKey='False'} (this
  can be the case for tables of type \tab{QuantityMatrix}, for instance).
\item \textbf{Order of columns} The allowed column types depend on the
  table type, but their order is arbitrary. The only exception is the
  first column, which contains the shortnames (acting as keys for this
  table) and whose name corresponds to the table type. However, it is
  good practice to sort the columns by importance and to arrange
  related columns next to each other (e.g. placing a column \col{Value} next to
  a column \col{Unit}).
\item \textbf{ASCII Characters} The table fields contain only plain text.
  \la{Special symbols like Greek letters are ignored.}  The format is
  case-sensitive, but the choice of fonts (bold, italic) does not play
  a role. Double quotes should not be used.
\item \textbf{Decimal points} To simplify parsing, we recommend to use
  decimal points (instead of decimal commas).
\item \textbf{Table types and column names} Table types and their
  possible columns are defined in appendix \ref{appendixA}.  Column
  names may not contain any special characters or white spaces
  (parsers should ignore these characters).
\item \textbf{Comment lines} Table lines starting with a ``\%''
  character contain comments and are ignored during parsing.
\item \textbf{Comments and references} Additional information about
  table elements can be  stored in the optional columns \col{!Comment},
  \col{!Reference}, \col{!Reference:Identifiers:pubmed}, and \col{!ReferenceDOI},
  which can appear in all tables.
\item \textbf{Unrecognised table or columns} Columns with unknown
  headers (not starting with \col{!}), or unrecognised header starting with
  \col{!} may appear in SBtab tables. They can be used, but
  are not supported by the parser. The use of undefined columns is inadviseable.
\item \textbf{Declaration line} The first line, starting with
  \tab{!!SBtab} must declare at least the attributes: \tab{TableType},
  \tab{TableName}, and possibly the properties \tab{SBtabVersion}
  (for SBtab version used) and \tab{Document}. The entries can be separated by whitespaces or
  be given in separate fields of the declaration line.
\item \textbf{Identifiers} Identifiers for compounds, compartments
  etc.~can be specified in columns with a header
  ``\emph{ElementType}:\col{Identifiers:}\emph{DB}'').  \la{\item
    \textbf{Qualifiers} SBtab supports the qualifiers (like ``Is'',
    ``VersionOf'') defined in the MIRIAM resources
    \cite{bioqualifiers}.  They can be used within table cells (in the
    syntax ``\emph{qualifier} \emph{Identifiers}''.}  \la{\item
    \textbf{Lists within table fields} Some table fields may contain
    several entries of the same type (e.g.~several alternative names
    for the same substance) separated by the ``$|$'' character.}
\item \textbf{Missing elements} If an element is missing, the table
  field is left empty. Missing numerical values can also be indicated by
  non-numerical elements like \texttt{?} or \texttt{na} (for ``not
  available''). Mandatory fields must not be empty.
\item \textbf{Formulae} Reaction sum formulae must be written in a
  special format explained below. \la{and formulae describing
    biochemical regulations}
\item \textbf{Reserved names} In the SBtab format, there are reserved
  names for (i) table types (marked by colours in this text); (ii)
  column names; (iii) types of biological elements (see Table
  \ref{tab:objects}); and (iv) types of biochemical quantities or
  mathematical terms (e.g.~\defint{Mean}) for them (see Table
  \ref{tab:quantities}), and physical units.
\item \textbf{Physical units} In SBtab, it is recommended to use the
  units listed in the SBML specification (see
  {sbml.org/Documents/Specifications})\footnote{The following units
    are supported by SBML: {\tt ampere, gram, katal, metre, second,
      watt, becquerel, gray, kelvin, mole, siemens, weber, candela,
      henry, kilogram, newton, sievert, coulomb, hertz, Litre, ohm,
      steradian, dimensionless, item, lumen, pascal, tesla, farad,
      joule, Lux, radian}. Orders of magnitude can be denoted by {\tt
      k, M, c, m, mu, n, p, f} for Kilo, Mega, Centi, Milli, Micro,
    Nano, Pico, Femto. If a parameter is dimensionless, it has to be
    annotated as {\tt dimensionless}.}. As good practice, derived
  units (e.g.  \defext{kJ/mol}) and reciprocal units
  (e.g. \defext{1/s}) should be given in the simplest possible form,
  in necessary using multiplication, division, exponentials, and round
  brackets (e.g.~\defext{gram/m\^\,3}).
\end{itemize}

\section{Overview of table types}
\label{appendixA}

\subsection{Document and table attributes and general column types}

\begin{table}[h!]
\textbf{Table attributes}\\[2mm]
\centerline{
{\footnotesize \begin{tabular}{|l|l|l|l|l|}
\hline
      \cellcolor{lightgrey} Name & 
      \cellcolor{lightgrey}Type & 
      \cellcolor{lightgrey}Format & 
      \cellcolor{lightgrey}Mandatory & 
      \cellcolor{lightgrey}Content \\
\hline
  \tab{TableType} & text & string & $\checkmark$ & Table type (as defined in definition table)\\
  \tab{TableName} & text & string & $\checkmark$ & Table name\\
 \tab{SBtabVersion} & text & string &      & SBtab version number \\
 \tab{Document} & text & string &          & SBtab document name \\
 \tab{UniqueKey} & text & Boolean &          & Requirement of a unique key identifier \\
  \hline
  \tab{ReferenceDescription} & text & string & & Name of reference description\\
  \tab{Document} & text & string & & Document name\\
  \tab{ReferenceCitation} & text & string & & Citation, unique identifier, unambiguous URL\\
  \tab{ModelCreators} & text & string &      & Name and contact information for model creators\\
 \tab{ModelCreationTime} & text & string &   & Date and time of model creation and last modification\\
 \tab{TermsOfDistribution} & text & string & & Terms of distribution \\
  \hline
\end{tabular}}}
\caption{Possible table attributes (to appear in declaration row). The
  attributes in the lower part would be necessary for MIRIAM
  compliance. If \tab{ReferenceCitation} contains a pubmed Id, the attribute 
  \tab{ReferenceCitation:Identifiers:pubmed} should be used instead.
\tab{ReferenceCitation} should also identify the authors of the model.}
\label{tab:columnsalltables}
\end{table}

\begin{table}[h!]
\textbf{All table types}\\[2mm]
{\footnotesize \begin{tabular}{|l|l|l|l|}
\hline
\cellcolor{lightgrey}Name & 
\cellcolor{lightgrey}Type & 
\cellcolor{lightgrey}Format & 
\cellcolor{lightgrey}Content \\
\hline
  \col{!Description} & text & string & Description of the row element\\
  \col{!Comment} & text & string & Comment\\
  \col{!ReferenceName} & text & string & Reference title, authors, etc.~(as free text)\\
  \col{!Reference:Identifiers:pubmed}& text & string & Reference PubMed ID\\
  \col{!ReferenceDOI} & text & string & Reference DOI\\
  \hline
\end{tabular}}
\caption{Columns that can appear in all tables}
\label{tab:columnsalltables}
\end{table}

\begin{table}[h!]
\textbf{All entity and reaction tables} \\[2mm]
{\footnotesize \begin{tabular}{|l|l|l|l|}
\hline
\cellcolor{lightgrey}Name & 
\cellcolor{lightgrey}Type & 
\cellcolor{lightgrey}Format & 
\cellcolor{lightgrey}Content \\
\hline
  \col{!Name} & text & string & Entity name \\
  \col{!Identifiers:}\emph{DataCollection} & resource ID & string & Entity ID \\
  \col{!MiriamAnnotations} & annotation & string  & Entity ID (JSON string)\\
  \col{!Type}	& text & string & Biochemical type of entity (examples see Table \ref{tab:objects}) \\
  \col{!Symbol} & text & string & Short symbol (e.g., gene symbol)\\
  \col{!PositionX} & number & float & x coordinate for graphical display \\
  \col{!PositionY} & number & float & y coordinate for graphical display \\
\hline
\end{tabular}}
\caption{Columns that can appear in all entity (i.e.~\tab{Compound},
  \tab{Enzyme}, \tab{Gene}, \tab{Regulator}, and \tab{Compartment})
  and \tab{Reaction} tables.}
\label{tab:columnsalltables}
\end{table}

\clearpage

\subsection{Predefined  table types}

\begin{table}[h!]
\tab{Compound}\\[2mm]
{\footnotesize \begin{tabular}{|l|l|l|l|}
\hline
\cellcolor{lightgrey}Name & 
\cellcolor{lightgrey}Type & 
\cellcolor{lightgrey}Format & 
\cellcolor{lightgrey}Content \\
\hline
  \col{!Compound} & shortname & string & Compound shortname\\
  \col{!SBML:species:id} & SBML element ID & string & SBML Species ID of the entity \\
  \col{!SBML:speciestype:id}	& SBML element ID & string & SBML SpeciesType ID of the entity \\
  \col{!InitialValue} & number & float & Initial amount or concentration \\
  \col{!Unit} & string & string & Unit for initial value \\
  \col{!Location} & shortname & string & Compartment for localised entities\\
  \col{!State} & shortname & string & State of the entity \\
  \col{!CompoundSumFormula} & text & string & Chemical sum formula \\
  \col{!StructureFormula} & text & string & Chemical structure formula \\
  \col{!Charge} & number & integer & Electrical charge number \\
  \col{!Mass}	& number & float & Molecular mass \\
  \col{!Unit} & text & string & Physical unit\\
  \col{!IsConstant} & Boolean & Boolean & Substance with fixed concentrations\\
  \col{!EnzymeRole} & shortname & string & Enzymatic activity \\
  \col{!RegulatorRole} & shortname & string & Regulatory activity \\
\hline
\end{tabular}}
\caption{Columns that can appear in \tab{Compound} tables}
\label{tab:columnsalltables}
\end{table}

\begin{table}[h!]
\tab{Enzyme}\\[2mm]
{\footnotesize \begin{tabular}{|l|l|l|l|}
  \hline
\cellcolor{lightgrey}  Name & 
\cellcolor{lightgrey}Type & 
\cellcolor{lightgrey}Format & 
\cellcolor{lightgrey}Content \\
  \hline
  \col{!Enzyme} & shortname & string & Enzyme shortname\\
  \col{!CatalysedReaction}	& shortname & string & Catalysed reaction \\
  \col{!KineticLaw:Name} & name & string & Rate law (name as in SBO) \\
  \col{!KineticLaw:Identifiers.obo.sbo} & shortname & string & Rate law SBO identifier \\
  \col{!Pathway} & text & string & Pathway name (free text)\\
  \col{!Gene}   & shortname & string & Gene coding for enzyme (shortname)\\
  \col{!Gene:Name} & string & string & Gene coding for enzyme (name)\\
  \col{!Gene:Symbol} & string & string & Gene coding for enzyme (short symbol)\\
  \hline
\end{tabular}}
\caption{Columns that can appear in \tab{Enzyme} tables}
\label{tab:columnsentities1}
\end{table}

\begin{table}[h!]
{\tab{Protein} \\[2mm]
{\footnotesize \begin{tabular}{|l|l|l|l|}
  \hline
\cellcolor{lightgrey}  Name & 
\cellcolor{lightgrey}Type & 
\cellcolor{lightgrey}Format & 
\cellcolor{lightgrey}Content \\
\hline
  \col{!Protein} & shortname & string & Protein shortname\\
  \col{!Name} & text & string & Protein name\\
  \col{!Symbol} & string & string  & Protein symbol\\
  \col{!Gene} & shortname & string & Gene shortname\\
  \col{!Gene:Name} & text & string & Gene name\\
  \col{!Gene:Symbol} & string & string  & Gene symbol\\
  \col{!Gene:LocusName} & string & string & Gene locus name\\
  \col{!Mass} & number  & number & Protein mass\\
  \col{!Size} & number  & number & Protein size\\
\hline
\end{tabular}}}
\caption{Columns that can appear in \tab{Protein} tables}
\label{tab:columnsalltables}
\end{table}

\begin{table}[h!]
{
  \tab{Gene} \\[2mm]
{\footnotesize \begin{tabular}{|l|l|l|l|}
  \hline
\cellcolor{lightgrey}  Name & 
\cellcolor{lightgrey}Type & 
\cellcolor{lightgrey}Format & 
\cellcolor{lightgrey}Content \\
\hline
  \col{!Gene} & shortname & string & Gene shortname\\
  \col{!Name} & text & string & Gene name\\
  \col{!Symbol} & string & string  & Gene symbol\\
  \col{!LocusName} & string & string & Gene locus name\\
  \col{!GeneProduct}& shortname & string & Gene product shortname\\
  \col{!GeneProduct:Name}& string & string & Gene product name \\
  \col{!GeneProduct:Symbol}& string & string & Gene product symbol \\
  \col{!GeneProduct:SBML:species:id} & SBML element ID & string & SBML ID of protein \\
  \col{!Operon} & shortname & string & Operon in which gene is located\\
\hline
\end{tabular}}}
\caption{Columns that can appear in \tab{Gene} tables}
\label{tab:columnsalltables}
\end{table}

\begin{table}[h!]
  \tab{Regulator} \\[2mm]
{\footnotesize \begin{tabular}{|l|l|l|l|}
 \hline
\cellcolor{lightgrey}  Name & 
\cellcolor{lightgrey}Type & 
\cellcolor{lightgrey}Format & 
\cellcolor{lightgrey}Content \\
  \hline
  \col{!Regulator} & shortname & string & Regulator shortname\\
  \col{!State} & shortname & string & State of the regulator \\
  \col{!TargetGene} & shortname & string & Target gene\\
  \col{!TargetOperon} & shortname & string & Target operon\\
  \col{!TargetPromoter} & shortname & string & Target promoter\\
\hline
\end{tabular}}
\caption{Columns that can appear in \tab{Regulator} tables}
\label{tab:columnsalltables}
\end{table}

\begin{table}[h!]
  \tab{Compartment} \\[2mm]
  {\footnotesize \begin{tabular}{|l|l|l|l|}
    \hline
\cellcolor{lightgrey}    Name & 
\cellcolor{lightgrey}Type & 
\cellcolor{lightgrey}Format & 
\cellcolor{lightgrey}Content \\
    \hline
    \col{!Compartment} & shortname & string & Compartment shortname\\
    \col{!Identifiers:obo.sbo} & shortname & string & Compartment SBO term\\
    \col{!SBML:compartment:id} & SBML element ID & string & SBML Compartment ID \\
    \col{!OuterCompartment} & shortname & string & Surrounding compartment (short) \\
    \col{!OuterCompartment:Name} & string & string & Surrounding compartment (name) \\
    \col{!OuterCompartment:SBML:compartment:id} & SBML element ID & string & Surrounding compartment \\
    \col{!Size} & number & float & Compartment size \\
    \col{!Unit} & text & string & Physical unit\\
    \hline
  \end{tabular}}
\caption{Columns that can appear in \tab{Compartment} tables}
\label{tab:columnsalltables}
\end{table}

\begin{table}[h!]
\label{tab:columnsentities}
\end{table}

\begin{table}[h!]
  \tab{Reaction} \\[2mm]
{\footnotesize \begin{tabular}{|l|l|l|l|}
  \hline
\cellcolor{lightgrey}  Name & 
\cellcolor{lightgrey}Type & 
\cellcolor{lightgrey}Format & 
\cellcolor{lightgrey}Content \\
  \hline
  \col{!Reaction} & shortname & string & Reaction shortname \\
  \col{!SBML:reaction:id} & SBML element ID & string & SBML Reaction ID \\
  \col{!SumFormula} & SumFormula formula & string & Reaction sum formula \\
  \hline
  \col{!Location} & shortname & string & Compartment for localised reaction\\
  \col{!Enzyme} & shortname & string & Enzyme catalysing the reaction \\
  \col{!Model} & text & string & Model(s) in which reaction is involved \\
  \col{!Pathway} & text & string & Pathway(s) in which reaction is involved \\
  \col{!SubreactionOf} & shortname & string & Mark as subreaction of a (lumped) reaction\\
  \hline
  \col{!IsComplete} & Boolean & Boolean & Reaction formula includes all cofactors etc.\\
  \col{!IsReversible} & Boolean & Boolean & Reaction should be treated as irreversible\\
  \col{!IsInEquilibrium} & Boolean & Boolean & Reaction approximately in equilibrium\\
  \col{!IsExchangeReaction} & Boolean & Boolean & Some reactants are left out \\
  \hline
  \col{!Flux} & number & float & Metabolic flux through the reaction \\
  \col{!IsNonEnzymatic} & Boolean & Boolean & Non-catalysed reaction \\
  \col{!KineticLaw:Name} & name & string & Rate law (name as in SBO) \\
  \col{!KineticLaw:Identifiers.obo.sbo} & shortname & string & Rate law SBO identifier \\
  \col{!Gene} & shortname & string & see table type \tab{Enzyme} \\
  \col{!Gene:Symbol} & string & string & see table type \tab{Enzyme} \\
  \col{!Operon} & shortname & string & see table type \tab{Gene}\\
  \hline
  \col{!Enzyme:Name} & string &string & Name of enzyme \\
  \col{!Enzyme:Identifiers:ec-code} & string & string & EC number of enzyme \\
  \col{!Enzyme:SBML:species:id} & SBML element ID &string & SBML ID of enzyme \\
  \col{!Enzyme:SBML:parameter:id} & SBML element ID &string & SBML ID of enzyme \\
  \col{!Enzyme:SBML:reaction:parameter:id} & SBML element ID &string & SBML ID of enzyme \\
  \col{!BuildReaction} & Boolean & Boolean & Includereaction in SBML model\\
  \col{!BuildEnzyme} & Boolean & Boolean & Include enzyme in SBML model\\
  \col{!BuildEnzymeProduction} & Boolean & Boolean & Describe enzyme production in SBML model\\
  \hline
\end{tabular}}
\caption{Columns that can appear in  \tab{Reaction} tables. The lower section lists, again, column types from Table \ref{tab:columnsentities}.}
\label{tab:columnsreactions}
\end{table}

\begin{table}[h!]
  \tab{Relation} \\[2mm]
{\footnotesize \begin{tabular}{|l|l|l|l|}
  \hline
\cellcolor{lightgrey}  Name & 
\cellcolor{lightgrey}Type & 
\cellcolor{lightgrey}Format & 
\cellcolor{lightgrey}Content \\
  \hline
  \col{!Relation} & shortname & string & Type of quantitative relationship\\
  \col{!From} & shortname & string & Element at beginning of arrow \\
  \col{!To} & shortname & string & Element at arrowhead\\
  \col{!IsSymmetric} & Boolean & Boolean & Flag indicating non-symmetric relationships \\
  \col{!Value}:\emph{QuantityType} & number & float & Numerical value assigned to the relationship \\
  \hline
\end{tabular}}
\caption{Columns that can appear in \tab{Relation} tables.}
\label{tab:columnsrelations}
\end{table}

\begin{table}[h!]
\tab{Quantity} \\[2mm]
{\footnotesize \begin{tabular}{|l|l|l|l|}
\hline
\cellcolor{lightgrey}  Name & 
\cellcolor{lightgrey}Type & 
\cellcolor{lightgrey}Format & 
\cellcolor{lightgrey}Content \\
  \hline
  \col{!Quantity} & shortname & string & Quantity / SBML parameter ID \\
  \col{!QuantityType} & shortname & string & Quantity type (e.g.~from SBO) \\
  \emph{ValueType} & ValueType & string & Mathematical Term from table \ref{tab:quantityvaluetype}\\
  \col{!SBML:parameter:id} & SBML element ID & string & Parameter ID in SBML file \\
  \col{!SBML:reaction:parameter:id} & SBML element ID & string & Parameter ID in SBML file \\
  \col{!Unit} & text & string & Physical unit\\
  \col{!Scale} & text & string & Scale (e.g.~logarithm, see Table \ref{tab:quantityvaluetype}) \\
  \col{!Provenance} & text & string & Name of data source (free text)\\
  \col{!Condition} & text & string & experimental condition name (free text)\\
  \col{!pH} & number & float & pH value in measurement\\
  \col{!Temperature} & number & float & Temperature in measurement\\
  \col{!Location} & shortname & string & Compartment (shortname)\\
  \col{!Location:Name} & string & string & Compartment (name)\\
  \col{!Location:SBML:compartment:id} & SBML element ID & string& SBML ID of compartment` \\
  \col{!Compound} & shortname & string & Related compound (shortname)\\
  \col{!Compound:Name} & string & string & Related compound (name)\\
  \col{!Compound:Identifiers:}\emph{DataCollection} & resource ID &string & Compound ID \\
  \col{!Compound:SBML:species:id} & SBML element ID & string & SBML ID of compound \\
  \col{!Reaction} & shortname & string & Related reaction (shortname)\\
  \col{!Reaction:Name} & string & string & Related reaction (name) \\
  \col{!Reaction:Identifiers:}\emph{DataCollection} & resource ID & string & Reaction ID \\
  \col{!Reaction:SBML:reaction:id} & SBML element ID & string & SBML ID of reaction \\
  \col{!Enzyme} & shortname & string & Related enzyme (shortname)\\
  \col{!Enzyme:Name} & string & string & Related enzyme (name) \\
  \col{!Enzyme:Identifiers:}\emph{DataCollection} & resource ID & string & Enzyme ID \\
  \col{!Enzyme:SBML:species:id} & SBML element ID & string & SBML ID of enzyme \\
  \col{!Enzyme:SBML:parameter:id} & SBML element ID & string& SBML ID of enzyme \\
  \col{!Enzyme:SBML:reaction:parameter:id} & SBML element ID & string& SBML ID of enzyme \\
  \col{!Protein} & shortname & string & Related enzyme (shortname)\\
  \col{!Protein:Name} & string & string & Related enzyme (name) \\
  \col{!Protein:Identifiers:}\emph{DataCollection} & resource ID & string & Protein ID \\
  \col{!Protein:SBML:species:id} & SBML element ID & string & SBML ID of enzyme \\
  \col{!Protein:SBML:parameter:id} & SBML element ID & string& SBML ID of enzyme \\
  \col{!Protein:SBML:reaction:parameter:id} & SBML element ID & string& SBML ID of enzyme \\
  \col{!Gene} & shortname & string & Related gene \\
  \col{!Organism} & shortname & string & Related organism \\
\hline
\end{tabular}}
\caption{Columns for numerical values and experimental conditions in tables of type
\tab{Quantity}.}
\label{tab:columnsquantity}
\end{table}

\begin{table}[h!]
  \tab{Definition} \\[2mm]
{\footnotesize \begin{tabular}{|l|l|l|l|}
  \hline
\cellcolor{lightgrey}Name & 
\cellcolor{lightgrey}Type & 
\cellcolor{lightgrey}Content \\
  \hline
  \col{!Component}     & component name  & Name of component (table, column, attribute to be defined)\\
  \col{!ComponentType} & \texttt{Table, Column, Attribute}  & Type of component\\
  \col{!IsPartOf}      & component name  & name of parent component\\
  \col{!Format}        & String & Format \\
  \col{!Description}   & Text & Free text description of component\\
  \hline
\end{tabular}}
\caption{Columns that can appear in \tab{Definition} tables.}
\label{tab:columnsrelations}
\end{table}

\clearpage

\section{Predefined terms and recommended controlled vocabularies}
\label{appendixB}

\begin{table}[h!]
  \begin{center}
    {\footnotesize \begin{tabular}{|l|l|l|l|}
      \hline
\cellcolor{lightgrey}      ValueType & 
\cellcolor{lightgrey}Type & 
\cellcolor{lightgrey}Format & 
\cellcolor{lightgrey}Meaning\\
      \hline
      \defint{Value} & number & float & Simple value \\
      \defint{Mean} & number & float & Algebraic mean \\
      \defint{Std} & number & float (positive) & Standard deviation \\
      \defint{Min} & number & float & Lower bound \\
      \defint{Max} & number & float & Upper bound \\
      \defint{Median}& number & float & Median \\
      \defint{GeometricMean}& number & float & Geometric mean \\
      \defint{Sign} & sign & \{+,-,0\} & Sign \\
      \defint{ProbDist} & Free text & string & Prob.~distribution \\
      \hline
    \end{tabular}} \hspace{5mm}
    {\footnotesize \begin{tabular}{|l|l|l|l|}
      \hline
\cellcolor{lightgrey}      Scale & 
\cellcolor{lightgrey}Meaning\\
      \hline
      \defint{Lin} & Linear scale (no transformation) \\
      \defint{Ln} & Natural logarithm \\
      \defint{Log2} & Dual logarithm \\
      \defint{Log10} & Decadic logarithm \\
      \hline
    \end{tabular}}  \end{center}
  \caption{Terms for mathematical quantities and mathematical scales  recommended for use in SBtab.
    Names of probability distributions can be, for instance,
    \texttt{Normal, Uniform, LogNormal}.}
\label{tab:quantityvaluetype}
\end{table}

\begin{table}[h!]
  \begin{center}
    {\footnotesize \begin{tabular}{|l|l|l|l|}
\hline
\cellcolor{lightgrey}  Database & 
\cellcolor{lightgrey}MIRIAM URN & 
\cellcolor{lightgrey}Contents & 
\cellcolor{lightgrey}URI \\
\hline
  SBO & \defext{obo.sbo} & Quantities, rate laws & www.ebi.ac.uk/sbo/ \\
  CheBI & \defext{obo.chebi} & Metabolites & www.ebi.ac.uk/chebi/ \\
  Enzyme nomenclature & \defext{ec-code} & Enzymes & www.ebi.ac.uk/IntEnz/ \\
  KEGG Compound & \defext{kegg.compound} & Compounds & www.genome.jp/KEGG/ \\
  KEGG Reaction & \defext{kegg.reaction} & Reactions & www.genome.jp/KEGG/ \\
  KEGG Orthology& \defext{kegg.orthology}& Genes & www.genome.jp/KEGG/ \\
  UniProt & \defext{uniprot} & Proteins  & www.uniprot.org/ \\
  SGD & \defext{sgd} & Yeast gene loci   & www.yeastgenome.org/ \\
  Gene Ontology & \defext{obo.go} & Compartments & www.geneontology.org/ \\
  Taxonomy & \defext{taxonomy} & Organisms & www.ncbi.nlm.nih.gov/Taxonomy/ \\
  SGD & \defext{sgd} & Yeast proteins & www.yeastgenome.org/ \\
\hline
\end{tabular}}
\end{center}
\caption{A selection of databases to be used in SBtab. For the
  complete list, see the MIRIAM resources
  \cite{laibe2007miriam}. \la{Additional resources can be defined in a
    \tab{AnnotationResource} table. }}
 \label{tab:databases}
\end{table}

\begin{table}[h!]
  \begin{center}
  {\footnotesize \begin{tabular}{|l|l|l|}
      \hline
\cellcolor{lightgrey}      \textbf{Physical entity types} & 
\cellcolor{lightgrey}\\ \hline
      \defext{protein complex }& SBO:0000297 \\
      \defext{messenger RNA }& SBO:0000278 \\
      \defext{ribonucleic acid }& SBO:0000250 \\
      \defext{deoxyribonucleic acid} & SBO:0000251 \\
      \defext{polypeptide chain }& SBO:0000252 \\
      \defext{polysaccharide }& SBO:0000249 \\
      \defext{metabolite }& SBO:0000299 \\
      \defext{macromolecular complex}& SBO:0000296 \\
\hline
    \end{tabular}}\hspace{5mm}
  {\footnotesize \begin{tabular}{|l|l|l|}
      \hline
\cellcolor{lightgrey}      \textbf{Compartments} & 
\cellcolor{lightgrey}\\ \hline
      \defext{cell } & GO:0005623 \\
      \defext{extracellular space } & GO:0005615 \\
      \defext{membrane } & GO:0001602 \\
      \defext{cytosol } & GO:0005829 \\
      \defext{nucleus } & GO:0005634 \\
      \defext{mitochondrion } & GO:0005739 \\
      \hline
    \end{tabular}}
  \end{center}
\caption{Examples of biochemical entity types (with Systems Biology Ontology identifiers \cite{cjkw:11}) and
cell compartments (with Gene Ontology identifiers \cite{GO}).}
\label{tab:objects}
\end{table}

\begin{table}[h!]
  \begin{center} 
    {\footnotesize \begin{tabular}{|l|l|l|l|l|}
      \hline
\cellcolor{lightgrey}      Name & 
\cellcolor{lightgrey}SBO term & 
\cellcolor{lightgrey}Default unit & 
\cellcolor{lightgrey}Entities \\
      \hline
      \defext{standard Gibbs energy of formation} & SBO:0000582 & kJ/mol & Compound\\
      \defext{standard Gibbs energy of reaction} & SBO:0000583 & kJ/mol & Compound \\
      \defext{equilibrium constant} & SBO:0000281 & variable & Reaction \\
      \defext{forward maximal velocity} & SBO:0000324 & mMol/s & Enzymatic reaction\\
      \defext{reverse maximal velocity} & SBO:0000325 & mMol/s& Enzymatic reaction\\
      \defext{substrate catalytic rate constant} & SBO:0000321 & 1/s & Enzymatic reaction\\
      \defext{product catalytic rate constant} & SBO:0000320 & 1/s & Enzymatic reaction\\
      \defext{Michaelis constant} & SBO:0000027 & mM & Enzyme, Compound\\
      \defext{inhibitory constant} & SBO:0000261 & mM & Enzyme, Compound\\
      \defext{activition constant} & SBO:0000363 & mM & Enzyme, Compound\\
      \defext{Hill constant} & SBO:0000190 & dimensionless & Compound, Reaction\\
      \defext{concentration} & SBO:0000196 & mM & Compound \\
      \defext{biochemical potential} & SBO:0000303 & kJ/mol & Compound \\
      \defext{standard biochemical potential} & SBO:0000463 & kJ/mol & Compound\\
      \hline
      \defext{rate of reaction (amount)} & SBO:0000615   &  M/s  &  Reaction \\
      \defext{rate of reaction (concentration)} & SBO:0000614   &  mM/s  & Reaction  \\
      \defext{Gibbs free energy of reaction} &  SBO:0000617  &  kJ/mol  & Reaction \\
      \defext{standard Gibbs free energy of formation} &  SBO:0000582  &  kJ/mol  & Compound \\
      \defext{standard Gibbs free energy of reaction} &  SBO:0000583  &  kJ/mol  & Compound \\
      \defext{transformed standard Gibbs free energy of reaction} &  SBO:0000620  &  kJ/mol  &  Reaction\\
      \defext{transformed standard Gibbs free energy of formation} &  SBO:0000621  & kJ/mol   &  Compound\\
      \defext{transformed Gibbs free energy of reaction} &  SBO:0000622  & kJ/mol   &  Reaction\\
      \defext{thermodynamic temperature} &  SBO:0000147  &  K  & Location (optional) \\
      \defext{ionic strength} &  SBO:0000623  & mM   & Location (optional) \\
      \defext{pH} & SBO:0000304 & dimensionless & Location  (optional) \\
      \hline
    \end{tabular}}
  \end{center}
  \caption{A selection of quantity types to be used in SBtab in table
    types \tab{Quantity}.  The unit of equilibrium constants depends
    on the reaction stoichiometry.  More quantities can be found in
    the Systems Biology Ontology \cite{cjkw:11}.}
\label{tab:quantities}
\end{table}

\section{A note on MIRIAM-compliant models}

The MIRIAM rules for computational models \cite{Novere:05} have been
established to guarantee that published models contain complete and
unambiguous information, and that results from the models can be
verified.  Note that MIRIAM-compliance also involves criteria that
cannot be ensured by the file structure alone, but are related to how
the model was made, and to the existence of a reference publication
(which may or may not exist for a given SBtab file).  (i) The encoded
model structure must reflect the biological processes described by the
reference description.  (ii) The model must be instantiable in a
simulation: all quantitative attributes must be defined, including
initial conditions.  (iii) When instantiated, the model must be able
to reproduce all results given in the reference description within an
epsilon (algorithms, round-up errors).

However, to allow users to satisfy some of the MIRIAM requirements,
SBtab contains document attributes for information that is mandatory
for MIRIAM-compliance. These must be given in the declaration line of
the SBtab document in question, or in the declaration lines of at
least one tables belonging to the document (i) ReferenceDescription
(ii) DocumentName (iii) ReferenceCitation (complete citation, unique
identifier, unambiguous URL). The citation should identify the authors
of the model.  (iv) ModelCreators (name and contact information for
model creators) (v) ModelCreationTime (The date and time of model
creation and last modification) (vi) TermsOfDistribution (link to a
precise statement about the terms of it's distribution).

\end{appendix}

\end{document}